\documentclass[pra,twocolumn,aps,longbibliography,10pt,floatfix]{revtex4-2}
\usepackage{graphicx}
\usepackage{amssymb, amsmath}
\usepackage{braket}
\usepackage{hyperref}
\usepackage[all]{hypcap}
\hypersetup{colorlinks=true, linkcolor=blue, citecolor=blue, filecolor=blue, urlcolor=blue}

\newcommand{\printtitle}{Characterizing the functional role of quantum coherence in energy transfer}

\newcommand{\ketbra}[3][-1mu]{%
  \lvert #2\rangle\mspace{#1}\langle #3\rvert
}

\begin{document}

\title{\printtitle} 

\author{Hallmann Óskar Gestsson}%
\email{hallmann.gestsson.20@ucl.ac.uk}
\affiliation{Department of Physics and Astronomy, University College London, London WC1E 6BT, United Kingdom}
\author{Alexandra Olaya-Castro}%
\email{a.olaya@ucl.ac.uk}
\affiliation{Department of Physics and Astronomy, University College London, London WC1E 6BT, United Kingdom}

\date{October 14, 2026}\label{key}

\begin{abstract}

    Quantum coherence has been suggested to play a role in excitation energy transfer in open quantum systems, yet a quantitative approach to assessing its influence on the transfer process is still missing. 
    Using Nakajima–Zwanzig projection operators, we derive a general memory kernel identity that enables us to characterize and quantify the impact of coherence in the eigenenergy basis on a generalized rate of energy transfer. 
    Applying our approach to the electronic dynamics of a dimer coupled to a structured phonon bath, we demonstrate how quantum coherence acts to modulate energy transfer. 
    
\end{abstract}

\maketitle

Understanding the influence of system--environment interactions in quantum system dynamics is an open problem of interest across a variety of fields, ranging from quantum information processing \cite{Goold2016Feb, Harrington2022Oct} to energy and charge transport in solid-state \cite{Gorman2018Mar, Bachtold2022Dec, Toffoletti2025Apr} and photochemical systems \cite{Scholes2017Mar, Schultz2024Nov}. 
In these contexts, it is now clear that structured and strong system--environment interactions result in non-trivial dissipation and coherence dynamics, highlighting the active role the environment can play in supporting quantum coherent processes \cite{Richards2012Jan, Kolli2012Nov, Chin2013Feb, O'Reilly2014Jan, Blau2018Mar, Jumper2018Jun, Arsenault2020Mar, Sohoni2024Nov, Lorenzoni2025Oct, Gustin2025Dec}. 
Electronic excitation energy transfer in photoexcited chemical and biomolecular systems provides a fundamental example in which system–environment interactions and their interplay with quantum coherence (in the eigenenergy basis) are important for energy transfer \cite{Scholes2011Oct, Romero2017Mar, Ma2019Feb, Zhu2024Apr, Jha2026Jan}.
Therefore, a key question is how to characterize such coherence--environment interplay and precisely quantify the impact of quantum coherence on the energy transfer process \cite{Chenu2015Apr}.

From a theoretical perspective, researchers have addressed this question by assessing the “quantumness” of the system’s state using, for instance, entanglement quantifiers \cite{Fassioli2010Aug, Sifain2021Oct}, coherence measures \cite{Li2012Nov, Baumgratz2014Sep}, or in terms of delocalization \cite{Levi2014Mar, Scholes2014Mar}, and attempting to correlate it with a figure of merit of the process of interest, such as the rate of energy transfer or quantum efficiency \cite{Olaya-Castro2008Aug, Rebentrost2009Jul}. 
Although insightful, these approaches do not provide a rigorous framework for establishing predictive and quantitative assessments of the impact of quantum coherence on such processes. 
Recent works have aimed to achieve such a rigorous quantification by framing the problem from a perspective of quantum resource theory \cite{Liebert2026Jun}, but it is not yet clear how such an approach can be experimentally validated.

In this Letter, we address this problem by proposing a theoretical approach that allows systematic quantification of the contribution of quantum coherence to generalized rates of population transfer between the energy eigenstates of a system. 
We do this by mapping the dynamics of an open quantum system (in any regime of system--environment interaction) to a generalized population master equation and decomposing the resulting memory kernel, and associated transfer or transition rates, into coherence-dependent and coherence-independent terms. 
This approach then allows us to isolate and quantify the contribution of the system--environment-induced coherence, and furthermore clearly separate it from the influence of any initial-state coherence in the basis of the system’s energy eigenstates (e.g., coherence created via ultrafast laser excitation) \cite{Brumer2018Jun, Li2023Jul, Li2025Nov}.

\begin{figure}[b]
    \centering
    \includegraphics[width=\linewidth]{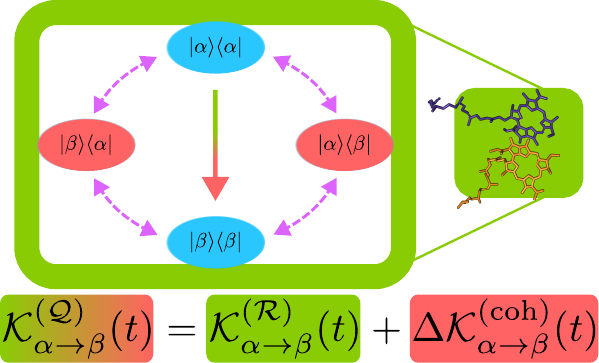}
    \caption{Electronic excitation dynamics in a model dimer of electronically coupled bacteriochlorophylls, with exciton eigenstates $\ket{\alpha}$ and $\ket{\beta}$, interacting strongly with phonon environments (green). 
    Environmental interactions induce population transfer between exciton states and couple populations to excitonic coherences. We introduce a projection-operator decomposition that separates the population transfer memory kernel into a thermal reference contribution $\mathcal{K}^{(\mathcal{R})}_{\alpha\to\beta}(t)$ and a contribution associated with coherence $\Delta\mathcal{K}_{\alpha\to\beta}^{(\textrm{coh})}(t)$, allowing us to quantify the influence of quantum coherence on energy transfer. }
    \label{fig:diagram}
\end{figure}

We illustrate the potential of our approach by considering an electronic dimer coupled to either an overdamped or an underdamped Brownian oscillator phonon environment and characterizing the influence of quantum coherence in modulating generalized excitonic energy transfer rates (see Fig.~\ref{fig:diagram}).~
Quantum coherence can therefore be clearly quantified as a mechanism that either directly enhances or suppresses energy transfer in comparison to thermal reference rates. 
Given that generalized transfer rates can in principle be extracted from ultrafast optical spectroscopic measurements \cite{Yuen-Zhou2014Jun, Tong2020Feb, Higgins2021Mar, Fleming2024Jan}, predictions based on our approach could be directly tested with standard ultrafast ensemble measurements.

Overall, our work advances the conceptual understanding of the impact of quantum coherence in energy transfer processes, with particular relevance to biological and photochemical settings \cite{Scholes2017Mar, Cao2020Apr}. 
The formalism is, however, general and versatile and can be applied to interrogate coherence contributions in dynamical processes of a wide range of open quantum systems.

\textit{Identification of quantum coherence contributions to energy transfer}—Our main question of interest is: assuming that a quantum system starts in a state with no coherence in the eigenenergy basis, does coherence contribute to the energy transfer? 
Addressing this question allows us to identify the impact of coherence that is driven by the system--environment interaction rather than coherence created in the initial state. 
This situation is of fundamental relevance for understanding the impact of coherence in natural and artificial systems when they are incoherently excited \cite{Fassioli2012Nov, Scholes2026Feb}.
In what follows, we outline the main steps to arrive at a kernel identity that allows us to formally isolate the impact of the environment-induced coherence on generalized energy transfer rates. 
The full derivation is presented in the Supplemental Material (SM) \cite{supplemental_material}.

We assume a separable system--environment initial state of the form $\hat{\rho}(0) = \hat{\rho}_S(0)\otimes\hat{\rho}_E$, where $\hat{\rho}_S(0)$ is the initial state of the system which is taken to be diagonal in the system's energy eigenbasis, i.e., there is no initial eigenstate coherence, and where $\hat{\rho}_{E}$ is the initial state of the environment. 
An exact treatment of the dynamics of the composite system and environment is formally determined by the Liouville–von Neumann equation $\partial_t\hat{\rho}(t) = \mathcal{L}\hat{\rho}(t)$, where $\hat{\rho}(t)$ is the joint system--environment state at time $t$. Often we are interested in observables of the system and in characterizing the influence of the system--environment interaction on the reduced system evolution which is obtained through a partial trace: $\hat{\rho}_S(t) = \text{Tr}_E(\hat{\rho}(t))$. To achieve a reduced system description, a standard technique is to introduce Nakajima--Zwanzig projection operators $\mathcal{P}$ and $\mathcal{Q} = \mathcal{I} - \mathcal{P}$, with $\mathcal{I}$ being the identity, while $\mathcal{P}$ and $\mathcal{Q}$ project on the “relevant” and the “irrelevant” parts of $\hat{\rho}(t)$ \cite{Nakajima1958Dec, Zwanzig1960Nov, Breuer2007Jan}. 
The relevant part is usually taken to be the reduced system density matrix $\hat{\rho}_S(t)$.

However, since we are interested in energy transfer dynamics, we may consider instead a modified $\mathcal{P}$ that projects onto the eigenstate population elements of the reduced system state as proposed in Refs.~\cite{Jesenko2013May, Zhang2016May}: denoting $\{\ket{\alpha}\}$ as the nondegenerate system energy eigenbasis, we set $\mathcal{P}\hat{\rho}(t) = \sum_\alpha p_\alpha(t)\ketbra{\alpha}{\alpha}\otimes\hat{\rho}_E$ with $p_\alpha(t) = \braket{\alpha\vert\hat{\rho}_S(t)\vert\alpha}$ being the population of eigenstate $\ket{\alpha}$. 
Then $\mathcal{Q}\hat{\rho}(t) = \hat{\rho}(t) - \mathcal{P}\hat{\rho}(t)$ will retain information on the environment dynamics, system--environment correlations, and any dynamically generated system coherence (given that the system starts in an initial state with no coherence).
An exact generalized master equation for the system's populations may then be obtained by tracing out the environment:
\begin{equation}\label{eq:bipartite_na-zw_eq}
    \partial_t p_\beta(t) = \sum_{\alpha} \int_0^t\text{d}\tau \mathcal{K}^{(\mathcal{Q})}_{\alpha\to\beta}(\tau) p_\alpha(t - \tau),
\end{equation}
with $\mathcal{K}^{(\mathcal{Q})}_{\alpha\to\beta}(t) = \braket{\beta\vert \text{Tr}_E(\mathcal{PL}e^{\mathcal{QL}t}\mathcal{QL}[\ketbra{\alpha}{\alpha}\otimes\hat{\rho}_E]) \vert\beta}$ being the $\ketbra{\alpha}{\alpha}\to\ketbra{\beta}{\beta}$ transition matrix elements. 
Equation (\ref{eq:bipartite_na-zw_eq}) follows from having assumed $\hat{\rho}(0) = \hat{\rho}_S(0)\otimes\hat{\rho}_E$ and $\mathcal{PLP} = 0$ (see SM for a derivation \cite{supplemental_material}). 
The memory kernel elements $\mathcal{K}^{(\mathcal{Q})}_{\alpha\to\beta}(t)$ capture the concurrent influence of both the environment and the dynamically generated coherences on population transfer as previously investigated \cite{Jesenko2013May, Zhang2016May}. 
However, since the impact of coherence is inextricably linked to environment influence, it cannot be resolved from $\mathcal{K}^{(\mathcal{Q})}_{\alpha\to\beta}(t)$ alone. 
To address this problem, we propose a simple but insightful expansion of $\mathcal{K}^{(\mathcal{Q})}_{\alpha\to\beta}(t)$ into two terms: one describing the environmental influence that excludes system coherences, and another contingent on the presence of such system quantum coherences.
We achieve this by decomposing the complementary projector as $\mathcal{Q} = \mathcal{C} + \mathcal{R}$, where $\mathcal{C}$ projects onto the reduced system eigenstate coherences, i.e., $\mathcal{C}\hat{\rho}(t) = \sum_{\alpha\neq\beta}\braket{\alpha\vert\hat{\rho}_S(t)\vert\beta}\ketbra{\alpha}{\beta}\otimes\hat{\rho}_E$, and $\mathcal{R}$ retains information on both the system--environment correlations and deviations of the environment from its initial reference state, i.e., $\mathcal{R}\hat{\rho}(t) = \hat{\rho}(t) - \hat{\rho}_S(t)\otimes\hat{\rho}_E$. 
Using the general operator identity $e^{(A+B)t} = e^{At} + \int_0^t\text{d}\tau e^{(A+B)\tau}Be^{A(t-\tau)}$ and setting $A=\mathcal{RL}$ and $B=\mathcal{CL}$, we can show that $\mathcal{K}^{(\mathcal{Q})}_{\alpha\to\beta}(t)$ expands as
\begin{equation}\label{eq:rate_decomposition}
    \mathcal{K}^{(\mathcal{Q})}_{\alpha\to\beta}(t) = \mathcal{K}^{(\mathcal{R})}_{\alpha\to\beta}(t) + \Delta\mathcal{K}_{\alpha\to\beta}^{(\textrm{coh})}(t),
\end{equation}
where $\mathcal{K}^{(\mathcal{R})}_{\alpha\to\beta}(t)$ and $\Delta\mathcal{K}_{\alpha\to\beta}^{(\textrm{coh})}(t)$ are the decomposed transition matrix elements after applying the operator identity (see SM for full details \cite{supplemental_material}). 
Our decomposition therefore resolves the population dynamics into two mechanisms: population transfer directly induced by the environment $\mathcal{K}^{(\mathcal{R})}_{\alpha\to\beta}(t)$, and population transfer mediated by environment-induced coherence $\Delta\mathcal{K}_{\alpha\to\beta}^{(\textrm{coh})}(t)$.
Equation~(\ref{eq:rate_decomposition}) constitutes an exact and general approach to characterizing the functional role of quantum coherence in energy transfer and is the main result of our Letter. 
It enables us to quantify the extent to which quantum coherence impacts the rate of an energy transfer process as we now discuss.

In the long-time regime, we can determine generalized population transfer rates as
\begin{equation}
    k_{\alpha\to\beta}^{(X)} = \int_0^\infty\text{d}t \mathcal{K}^{(X)}_{\alpha\to\beta}(t),
\end{equation}
for $\alpha\neq\beta$ and $X=\mathcal{Q},\mathcal{R}$. 
We refer to $k^{(\mathcal{Q})}_{\alpha\to\beta}$ as the total transfer rate and $k^{(\mathcal{R})}_{\alpha\to\beta}$ as the reference rate that is independent of coherences. 
Hence, from Eq.~(\ref{eq:rate_decomposition}) we conclude that $k^{(\mathcal{Q})}_{\alpha\to\beta} - k^{(\mathcal{R})}_{\alpha\to\beta}$ quantifies the contribution of coherences to the population transfer process. 
We therefore consider as a figure of merit the following dimensionless ratio
\begin{equation}\label{eq:coherent_contributions}
    f_{\alpha\to\beta} = 1 - \frac{k^{(\mathcal{R})}_{\alpha\to\beta}}{k^{(\mathcal{Q})}_{\alpha\to\beta}}
\end{equation}
that quantifies the relative contribution of environment-induced coherence to the total transfer rate for a given $\alpha\to\beta$ process. 
Assuming $k_{\alpha\to\beta}^{(\mathcal{Q})}$, $k_{\alpha\to\beta}^{(\mathcal{R})} > 0$, a positive or negative $f_{\alpha\to\beta}$ is an indication of quantum coherence acting to either enhance or suppress the transfer process, respectively, whereas $f_{\alpha\to\beta} = 0$ implies either that there is no environment-induced quantum coherence or that the influence of quantum coherence averages itself out. 
Therefore, $f_{\alpha\to\beta}\neq 0$ not only witnesses environment-induced coherence but also characterizes its functional role in population transfer.

\begin{figure}
    \centering
    \includegraphics[width=\linewidth]{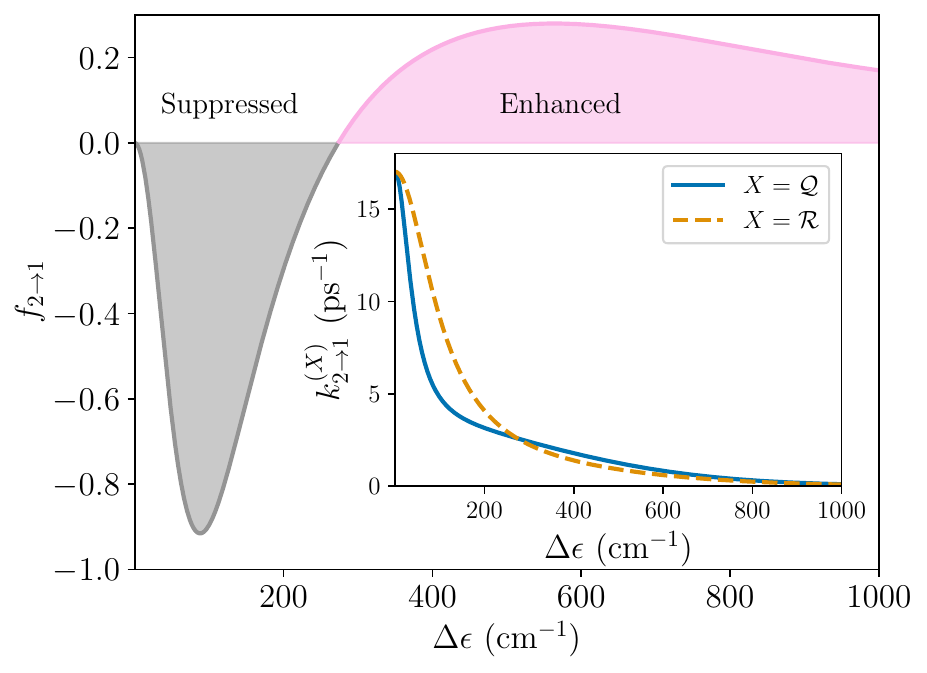}
    \caption{The figure of merit $f_{2\to1}$ as a function of detuning in a dimer coupled to an overdamped Brownian oscillator. 
    The gray and pink regions correspond to regimes where the functional role of quantum coherence is to suppress and enhance the energy transfer process, respectively. 
    The inset shows the corresponding total rate $k^{(\mathcal{Q})}_{2\to1}$ (solid line) and thermal reference rate $k^{(\mathcal{R})}_{2\to1}$ (dashed line) of the downhill energy transfer process. 
    The two rates deviate for intermediate detuning, demonstrating the impact of quantum coherence on the transfer process. }
    \label{fig:rates}
\end{figure}

\textit{Electronic excitation energy transfer in a dimer}—As an illustrative example of the impact that environment-induced quantum coherence has on energy transfer, we consider a dimer model consisting of two sub-units that are electronically coupled and interacting with identical local thermal phonon environments. 
We denote the local excited states of each sub-unit as $\ket{e_1}$ and $\ket{e_2}$, such that the electronic component of the Hamiltonian with detuning $\Delta\epsilon$ and coupling strength $V$ can be written as $\hat{H}_S = \frac{\Delta\epsilon}{2}(\ketbra{e_1}{e_1} - \ketbra{e_2}{e_2}) + V(\ketbra{e_1}{e_2} + \ketbra{e_2}{e_1})$ (we have set $\hbar=1$ throughout the Letter). 
The exciton eigenstates of $\hat{H}_S$ are denoted as $\ket{1}$ and $\ket{2}$, and they are separated in energy by the exciton energy gap $\Delta E = \sqrt{\Delta\epsilon^2 + 4V^2}$. 
Population transfer between excitonic states of the dimer occurs then due to a combination of thermal relaxation ($\mathcal{K}^{(\mathcal{R})}_{\alpha\to\beta}(t)$) and excitonic coherence ($\Delta\mathcal{K}_{\alpha\to\beta}^{(\textrm{coh})}(t)$). 
The influence of such coherence on population transfer rates may now be quantified precisely according to Eq.~(\ref{eq:coherent_contributions}), enabling a characterization of its functional role in the excitonic energy transfer process. 
The phonon environments are modeled by two identical ensembles of normal modes: $\hat{H}_E = \sum_{i,k}\omega_k\hat{b}_{i,k}^\dagger\hat{b}_{i,k}$ where $\hat{b}_{i,k}^\dagger$/$\hat{b}_{i,k}$ are bosonic creation/annihilation operators satisfying the commutation relation $[\hat{b}_{i,k},\hat{b}_{j,k^\prime}^\dagger] = \delta_{i,j}\delta_{k,k^\prime}$.
The local electronic--vibrational interaction is taken to be of the form $\hat{H}_I = \sum_{i,k}g_k\ketbra{e_i}{e_i}\otimes(\hat{b}_{i,k}^\dagger + \hat{b}_{i,k})$, with a spectral density of the Drude–Lorentz form $J(\omega) = 2\lambda\gamma\omega / (\omega^2 + \gamma^2)$. 
The electronic coupling is $V = 50$~cm$^{-1}$ and the $J(\omega)$ parameters are $\lambda = 100$~cm$^{-1}$, $\gamma = 53$~cm$^{-1}$, and we assume a room temperature of $T = 300$~K (environmental parameters are chosen to coincide with Refs.~\cite{Yang2002Jan, Ishizaki2009Jun}). 
We assume each local phonon environment is initially in a thermal state, i.e., $\hat{\rho}_E = e^{-\hat{H}_E / k_\text{B}T} / \text{Tr}(e^{-\hat{H}_E / k_\text{B}T})$, where $k_\text{B}$ is the Boltzmann constant. 
Further details for the dimer model are discussed in the SM \cite{supplemental_material}. 
For this case the projection operators are $\mathcal{P}\hat{\rho}(t) = (\braket{1\vert\hat{\rho}_S(t)\vert 1}\ketbra{1}{1} + \braket{2\vert\hat{\rho}_S(t)\vert 2}\ketbra{2}{2})\otimes\hat{\rho}_E$ and $\mathcal{C}\hat{\rho}(t) = (\braket{1\vert\hat{\rho}_S(t)\vert 2}\ketbra{1}{2} + \braket{2\vert\hat{\rho}_S(t)\vert 1}\ketbra{2}{1})\otimes\hat{\rho}_E$, with $\mathcal{R}\hat{\rho}(t)$ containing information on the system--environment correlations as well as deviations of the local environments from thermal equilibrium.
We assume $\hat{\rho}(0)=\hat{\rho}_S(0)\otimes \hat{\rho}_E$ where $\hat{\rho}_S(0) = p_1\ketbra{1}{1} + p_2\ketbra{2}{2}$ with $0\leq p_1,p_2\leq1$ and $p_1 + p_2 = 1$, i.e., an incoherent initial excitonic state.
With $\ket{2}$ as the highest energy exciton eigenstate, the figure of merit $f_{2\to1}$ unequivocally quantifies and characterizes the impact of dynamically generated excitonic quantum coherence on exciton population transfer.
To compute $f_{2\to1}$ we have applied the hierarchical equations of motion which describe, in a non-perturbative manner, the influence of phonon environments on the reduced system dynamics \cite{Tanimura1989Jan, Ishizaki2005Dec, Shi2009Feb, Ishizaki2009Jun, Tanimura2020Jul}. 
The SM gives the full details of our numerical implementation \cite{supplemental_material}.

Figure~\ref{fig:rates} shows $f_{2\to1}$ as a function of the dimer detuning $\Delta\epsilon$, with the inset depicting the rates $k^{(\mathcal{Q})}_{2\to1}$ and $k^{(\mathcal{R})}_{2\to1}$.
Results for the uphill process, $k_{1\to2}^{(X)}$, and the deviations from the bare detailed balance ($k_{2\to1}^{(X)}/k_{1\to2}^{(X)} \neq e^{\Delta E / k_\text{B}T}$), as expected for strong system--environment interaction, can be found in the SM \cite{supplemental_material}.
We find $f_{2\to1}$ vanishes when $\Delta\epsilon = 0$, $\Delta\epsilon\to\infty$, and at the crossover point where the impact of quantum coherence transitions from negative to positive. 
The crossover occurs when the exciton energy gap $\Delta E$ roughly coincides with the thermal energy scale, $k_\text{B}T = 208$~cm$^{-1}$. 
In the limit of very large $\Delta\epsilon\approx1000$~cm$^{-1}$, the transfer rate is significantly reduced, as expected, yet the coherence contribution remains non-negligible. 
This analysis reveals the functional impact of environment-induced exciton coherence: a significant reduction of the transfer rate with respect to the thermal reference in the effective high-temperature case ($\Delta E \ll k_\text{B}T$) and a modest increase of the rate in the effective low-temperature case ($\Delta E \gg k_\text{B}T$). 
Interestingly, several photosynthetic complexes have pairs of chromophores that facilitate interband transfer and operate in the regime where $\Delta E\sim \Delta\epsilon > k_\text{B}T > V$, which indicates their potential for exploiting environment-induced coherence \cite{O'Reilly2014Jan, Scott2025Nov}.

An important case to consider is when $\Delta\epsilon = 0$ and $V \neq 0$, where excitonic eigenstates are maximally delocalized.
For this resonant scenario, we find that the transfer rate $k^{(\mathcal{Q})}_{2\to1}$ is the largest, and it coincides with the thermal reference $k^{(\mathcal{R})}_{2\to1}$ yielding $f_{2\to1} = 0$. 
This means that quantum coherence does not contribute to the population transfer process. 
The origin of this result is the exchange symmetry with respect to the two sites, with the eigenstates $\ket{2}$ and $\ket{1}$ having even and odd parity, respectively.
For the initial incoherent diagonal exciton state assumed here, the system–environment dynamics preserves this exchange symmetry, and therefore it cannot generate coherence between excitonic states of opposite parity. Consequently, for $\Delta\epsilon=0$, excitonic coherence generation by the environment is forbidden for any nonzero $V$ and any $J(\omega)$ (see proof in SM \cite{supplemental_material}). 
Hence, $\Delta\mathcal{K}_{\alpha\to\beta}^{(\textrm{coh})}(t) = 0$ for $t\geq0$, which leads to $k^{(\mathcal{Q})}_{2\to1}$=$k^{(\mathcal{R})}_{2\to1}$ and therefore $f_{2\to1} = 0$. 
This result suggests an important design principle: quantum coherence contributions to an energy transfer process can be activated and tuned by modulating the electronic energy detuning of the system, as shown in Fig.~\ref{fig:rates}. 
It is notable that natural light-harvesting antennae exhibit mechanisms, such as static disorder, that break such symmetries and modify energy detunings, suggesting they may exploit such symmetry-breaking principles to exploit environment-induced coherence.

\begin{figure}
    \centering
    \includegraphics[width=\linewidth]{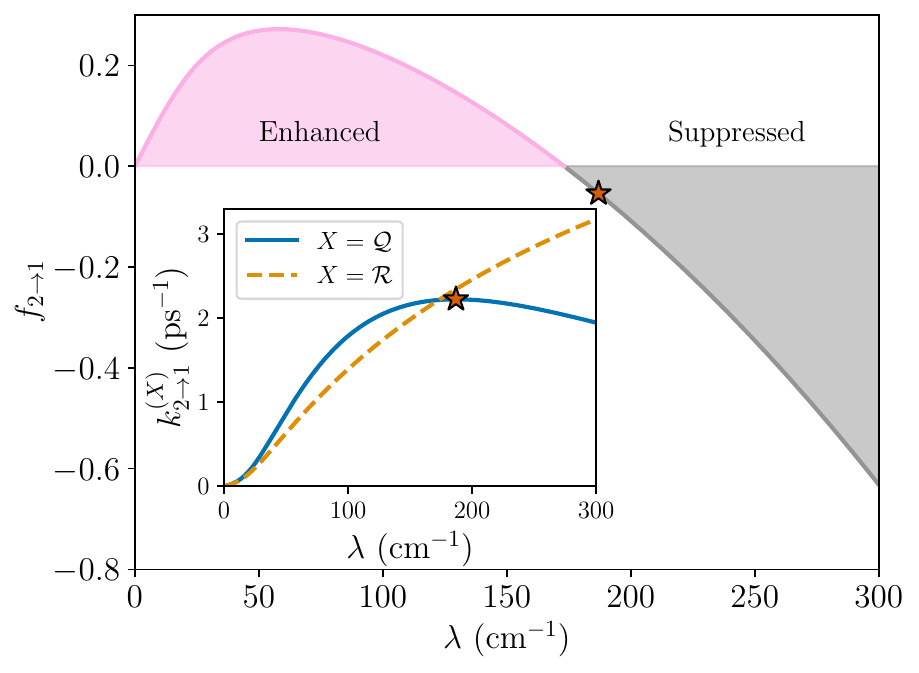}
    \caption{The figure of merit $f_{2\to1}$ as a function of the reorganization energy of the overdamped Brownian oscillator, with $\Delta\epsilon = 400$~cm$^{-1}$. 
    The gray and pink regions correspond to regimes where the functional role of quantum coherence is to suppress and enhance the energy transfer process, respectively.
    The inset shows the corresponding total rate $k^{(\mathcal{Q})}_{2\to1}$ (solid line) and thermal reference rate $k^{(\mathcal{R})}_{2\to1}$ (dashed line) of the downhill transfer process. 
    The star marker indicates the point where the optimal value of $k^{(\mathcal{Q})}_{2\to1}$ is achieved, which in this example occurs with associated $f_{2\to1}<0$. }
    \label{fig:rates_v_reorg}
\end{figure}

For the results shown in Fig.~\ref{fig:rates_v_reorg}, we have fixed $\Delta\epsilon = 400$~cm$^{-1}$ and $V = 50$~cm$^{-1}$ while varying the reorganization energy $\lambda$, thereby depicting the behavior of $f_{2\to1}$ going from weak to strong system–environment interaction, with the inset showing the corresponding $k^{(\mathcal{Q})}_{2\to1}$ and $k^{(\mathcal{R})}_{2\to1}$. 
Previous works indicate that one should expect an optimal value of $\lambda$ for which $k^{(\mathcal{Q})}_{2\to1}$ is maximal \cite{Ishizaki2009Jun}, which is what we observe here as well. 
However, the maxima of $f_{2\to1}$ and $k^{(\mathcal{Q})}_{2\to1}$ do not coincide, and we find $f_{2\to1}<0$ where the maximum of $k^{(\mathcal{Q})}_{2\to1}$ is achieved, as indicated by the star marker in Fig.~\ref{fig:rates_v_reorg}. 
Thus, in this example, energy transfer is optimized when the effect of coherence is to suppress, rather than enhance, transfer.
In the SM, we present analogous examples with different $V$, for which the maximum rate can occur when coherence enhances transfer, suppresses transfer, or does neither \cite{supplemental_material}. 
Therefore, our results demonstrate that coherence does indeed play a role in achieving such optimal behavior, not because it enhances the rate, but because it modulates the otherwise monotonic growth of $k^{(\mathcal{R})}_{2\to1}$ as a function of $\lambda$. 
This analysis highlights the importance of $f_{2\to1}$ in unveiling the underlying influence of coherence.

\begin{figure}
    \centering
    \includegraphics[width=\linewidth]{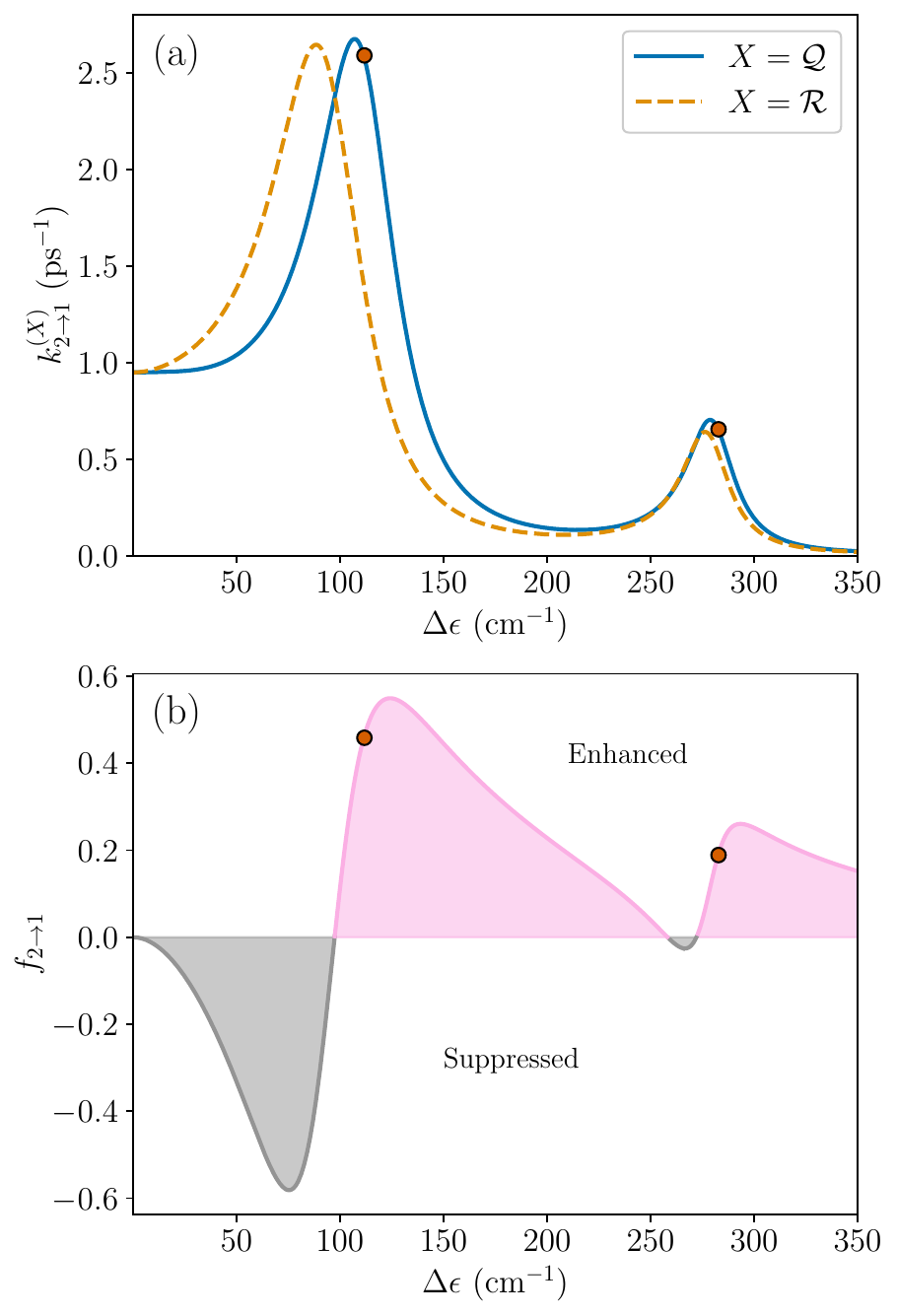}
    \caption{(a) The total rate $k^{(\mathcal{Q})}_{2\to1}$ (solid line) and thermal reference rate $k^{(\mathcal{R})}_{2\to1}$ (dashed line) as a function of detuning in the dimer coupled to an underdamped Brownian oscillator. 
    The mode is centered at $\Omega = 150$~cm$^{-1}$, such that exciton--vibration resonances occur at detuning values $\sqrt{n^2\Omega^2 - 4V^2}$ for integers $n=1,2,\ldots$, which are denoted by the dot markers. 
    (b) Corresponding figure of merit $f_{2\to1}$ as a function of $\Delta\epsilon$. 
    The gray and pink regions correspond to regimes where the functional role of quantum coherence is to suppress and enhance the energy transfer process, respectively. 
    }
    \label{fig:rates_with_ubo}
\end{figure}

Let us now consider the dimer's local electronic states to each be coupled to a broadened vibrational mode with a linewidth $\gamma$ which we model using an underdamped Brownian oscillator spectral density: $J(\omega) = 2\lambda\gamma\Omega^2\omega / ((\Omega^2 - \omega^2)^2 + \gamma^2\omega^2)$. 
The electronic coupling is again $V = 50$~cm$^{-1}$ and the spectral density parameters are $\lambda = 10$~cm$^{-1}$, $\gamma = 5$~cm$^{-1}$, and $\Omega = 150$~cm$^{-1}$, assuming a room temperature of $T = 300$~K. 
The functional role of quantum coherence and its potential to be tuned via exciton–phonon resonance conditions to modulate energy transfer have recently been suggested as a photoprotective mechanism exploited by light-harvesting antenna complexes \cite{Higgins2021Mar, Gestsson2025Mar}.

Figure~\ref{fig:rates_with_ubo} depicts $k_{2\to 1}^{(\mathcal{Q})}$, $k_{2\to 1}^{(\mathcal{R})}$, and $f_{2\to 1}$ as functions of $\Delta\epsilon$, illustrating the impact of exciton--phonon resonance on the excitonic energy transfer rates (Fig.~\hyperref[fig:rates_with_ubo]{4(a)}), and the functional role of quantum coherence therein (Fig.~\hyperref[fig:rates_with_ubo]{4(b)}). 
We again find $f_{2\to1} = 0$ for $\Delta\epsilon = 0$ and $\Delta\epsilon\to\infty$, as these results are independent of the specific form of $J(\omega)$. 
Figure~\hyperref[fig:rates_with_ubo]{4(a)} demonstrates the rich behavior of $k_{2\to 1}^{(\mathcal{Q})}$ and $k_{2\to 1}^{(\mathcal{R})}$ when the detuning is varied across the single- and double-phonon resonance conditions that promote the energy transfer process, as indicated by the occurrence of local maxima in the rates of transfer. 
The role of environment-induced coherence in the downhill transfer process is revealed by the corresponding $f_{2\to1}$ results shown in Fig.~\hyperref[fig:rates_with_ubo]{4(b)}. 
We observe a suppressive influence of quantum coherence for $\Delta\epsilon$ values below the single-phonon resonance which then quickly transitions to being an enhancing influence at resonance, and similarly to a lesser extent around the double-phonon resonance. 
Our figure of merit $f_{2\to1}$ therefore enables a transparent and direct assessment of the influence of quantum coherence for systems interacting with complicated structured environments such as the underdamped Brownian oscillator. 
A natural light-harvesting antenna will have a spectral density that is more complex than the single mode we have considered here, but our approach is nonetheless applicable to such scenarios. 
We have limited our analysis to the most illustrative examples of structured environments to clearly highlight the influence of environment-induced coherence on energy transfer.

\textit{Conclusions and outlook}—We have devised a theoretical approach to quantify and characterize the influence of quantum coherence in energy transfer processes. Our main focus has been the impact of system coherence in the eigenstate basis that is generated dynamically through system–environment interactions, in the absence of initial-state coherence. This advances our understanding of quantum effects in the dynamics of photoexcited physical, biomolecular and chemical systems under incoherent excitation. The power of this approach is illustrated by investigating energy transfer in a prototype light-harvesting electronic dimer interacting strongly with phonon environments. We demonstrated that environment-induced excitonic quantum coherence plays an important role in modulating energy transfer processes as a function of the system–environment coupling strength: depending on the regime of operation, quantum coherence contributes to either enhance or suppress generalized energy transfer rates in a quantifiable manner. 
Our analysis also shows that a nonzero excitation energy detuning, whereby the exchange symmetry in the dimer is broken, is necessary for such environment-induced quantum coherence to play any role at all, suggesting a design principle for light-harvesting systems exploiting environment-induced coherence.
While our focus is on initial states with no coherence, the general derivation presented in the SM shows that these initial-state assumptions can be relaxed to quantify and distinguish the influence of coherence that may be initially prepared. Our approach can therefore be applied to larger natural or synthetic multichromophoric systems to investigate the functional role of quantum coherence, whether induced by the environment, prepared in the initial state or sustained under non-equilibrium steady-state conditions \cite{Brumer2018Jun}.

The proposed methodology can be used in conjunction with a variety of non-perturbative treatments of open quantum dynamics and it can be extended to quantify the influence of mechanisms other than coherence.
For instance, by decomposing the projection on the phonon environment in terms of subsets of degrees of freedom (e.g., particular vibrational modes accessible in an experiment), the contribution of such a subset to generalized population transfer rates can be quantified.
Finally, the approach may be extended to understand the impact of coherence on other dynamical processes such as thermodynamic work \cite{Francica2019Apr, Hammam2022Dec, Rodrigues2024Feb}, thermokinetic uncertainty relations \cite{Prech2025Jan}, and quantum speed limits \cite{delCampo2026May}.

\textit{Acknowledgments}—H.Ó.G. and A.O.-C. acknowledge funding from the Gordon and Betty Moore Foundation (GBMF 8820) and the Engineering and Physical Sciences Research Council (EPSRC) (UKRI2297).

\textit{Data availability}—The data that support the findings of this article are not publicly available. The data are available from the authors upon reasonable request.

\let\savedaddcontentsline\addcontentsline
\renewcommand{\addcontentsline}[3]{}

\bibliography{bib}

\let\addcontentsline\savedaddcontentsline

\onecolumngrid
\clearpage

\begin{center}
    \textbf{\large Supplemental Material for: \\ “\printtitle”}
\end{center}
\begin{center}
    {\normalsize Hallmann Óskar Gestsson and Alexandra Olaya-Castro}
    
    {\small\itshape Department of Physics and Astronomy, University College London, London WC1E 6BT, United Kingdom}
    
    {\small (Dated: \today)}
\end{center}

\setcounter{equation}{0}
\setcounter{figure}{0}
\setcounter{page}{1}
\makeatletter
\renewcommand{\theequation}{S\arabic{equation}}
\renewcommand{\thefigure}{S\arabic{figure}}
\renewcommand{\thepage}{S\arabic{page}}
\renewcommand{\theHequation}{Supplement.\theequation}
\renewcommand{\theHfigure}{Supplement.\thefigure}

\thispagestyle{empty}

\renewcommand{\tocname}{Table of contents}
\tableofcontents

\section{Nakajima–Zwanzig equation of motion derivation and decomposition of the total coherence--environment influence}
The dynamics of the joint system–environment density matrix $\hat{\rho}(t)$ is given by the Liouville–von Neumann equation of motion: $\partial_t\hat{\rho}(t) = \mathcal{L}\hat{\rho}(t)$. 
The reduced density matrix dynamics are recovered by a partial trace over the environment: $\hat{\rho}_S(t) = \text{Tr}_E(\hat{\rho}(t))$. 
Let us now introduce a projection operator onto a set of orthonormal basis state populations of our reduced system:
\begin{equation}
    \mathcal{P}\hat{\rho}(t) = \sum_{\alpha}\braket{\alpha\vert\hat{\rho}_S(t)\vert\alpha}\ketbra{\alpha}{\alpha}\otimes\hat{\rho}_E,
\end{equation}
where $\hat{\rho}_E$ is a static environmental reference state with $\text{Tr}(\hat{\rho}_E) = 1$. 
In the main text we have assumed $\ket{\alpha}$ to be the eigenstates of $\hat{H}_S$, an assumption that we will introduce later on in this section. 
The complementary projection operator is given by $\mathcal{Q} = \mathcal{I} - \mathcal{P}$, where $\mathcal{I}$ is the identity projection. 
We therefore have
\begin{align}
    \mathcal{Q}\hat{\rho}(t) &= \hat{\rho}(t) - \sum_{\alpha}\braket{\alpha\vert\hat{\rho}_S(t)\vert\alpha}\ketbra{\alpha}{\alpha}\otimes\hat{\rho}_E \\
    &= \hat{\rho}(t) - \hat{\rho}_S(t)\otimes\hat{\rho}_E + \sum_{\alpha\neq\beta}\braket{\alpha\vert\hat{\rho}_S(t)\vert\beta}\ketbra{\alpha}{\beta}\otimes\hat{\rho}_E,
\end{align}
such that we can see that $\mathcal{Q}$ projects onto both the so-called “irrelevant part” and the reduced system coherences, which we denoted in the main text as $\mathcal{R}\hat{\rho}(t) = \hat{\rho}(t) - \hat{\rho}_S(t)\otimes\hat{\rho}_E$ and $\mathcal{C}\hat{\rho}(t) = \sum_{\alpha\neq\beta}\braket{\alpha\vert\hat{\rho}_S(t)\vert\beta}\ketbra{\alpha}{\beta}\otimes\hat{\rho}_E$, respectively. 
The “irrelevant part” $\mathcal{R}\hat{\rho}(t)$ describes the system–environment correlations and the deviation of the environment from its initial state, information that is typically considered to be inaccessible in an experiment, such that an effective description of its influence on the reduced system is considered instead.

Following the standard Nakajima–Zwanzig projection operator derivation of the equation of motion for $\mathcal{P}\hat{\rho}(t)$ \cite{Breuer2007Jan}, we first note that $\partial_t\mathcal{P}\hat{\rho}(t) = \mathcal{P}\mathcal{L}\hat{\rho}(t)$ and $\partial_t\mathcal{Q}\hat{\rho}(t) = \mathcal{Q}\mathcal{L}\hat{\rho}(t)$. 
Next, by utilizing the fact that $\mathcal{I} = \mathcal{P} + \mathcal{Q}$, we have the closed set of equations 
\begin{align}
    \partial_t\mathcal{P}\hat{\rho}(t) &= \mathcal{P}\mathcal{L}\mathcal{P}\hat{\rho}(t) + \mathcal{P}\mathcal{L}\mathcal{Q}\hat{\rho}(t), \\
    \partial_t\mathcal{Q}\hat{\rho}(t) &= \mathcal{Q}\mathcal{L}\mathcal{P}\hat{\rho}(t) + \mathcal{Q}\mathcal{L}\mathcal{Q}\hat{\rho}(t).
\end{align}
We may formally solve for $\mathcal{Q}\hat{\rho}(t)$ as
\begin{equation}
    \mathcal{Q}\hat{\rho}(t) = e^{\mathcal{QL}t}\mathcal{Q}\hat{\rho}(0) + \int_0^t\text{d}\tau e^{\mathcal{QL}\tau}\mathcal{QLP}\hat{\rho}(t-\tau),
\end{equation}
where $\hat{\rho}(0)$ is the initial state of the joint system--environment, which is then inserted in the equation of motion for $\mathcal{P}\hat{\rho}(t)$ to obtain
\begin{equation}\label{eq:bi_naka_zwa}
    \partial_t\mathcal{P}\hat{\rho}(t) = \mathcal{PLP}\hat{\rho}(t) + \mathcal{PL}e^{\mathcal{QL}t}\mathcal{Q}\hat{\rho}(0) + \int_0^t\text{d}\tau \mathcal{K}_\mathcal{Q}(\tau)\mathcal{P}\hat{\rho}(t-\tau),
\end{equation}
where we have set $\mathcal{K}_\mathcal{Q}(\tau) = \mathcal{PL}e^{\mathcal{QL}\tau}\mathcal{QL}$ as the total memory kernel that describes the simultaneous influence of system--environment correlations and the reduced system coherences on the reduced system population dynamics.

Let us now make assumptions on the form of the initial state of the joint system--environment density matrix $\hat{\rho}(0)$. 
First, we make the standard factorization assumption, such that we have 
\begin{equation}
    \hat{\rho}(0) = \hat{\rho}_S(0)\otimes\hat{\rho}_E. 
\end{equation}
This directly leads to $\mathcal{R}\hat{\rho}(0) = 0$ and $\mathcal{Q}\hat{\rho}(0) = \mathcal{C}\hat{\rho}(0)$. 
Second, we assume that the initial state of the reduced system is diagonal in the $\ket{\alpha}$ basis such that 
\begin{equation}
    \hat{\rho}_S(0) = \sum_\alpha \braket{\alpha\vert\hat{\rho}_S(0)\vert\alpha}\ketbra{\alpha}{\alpha},
\end{equation}
leading to $\mathcal{C}\hat{\rho}(0) = 0$. 
Equation (\ref{eq:bi_naka_zwa}) therefore reduces to
\begin{equation}\label{eq:generalized_master_equation_general}
    \partial_t\mathcal{P}\hat{\rho}(t) = \mathcal{PLP}\hat{\rho}(t) + \int_0^t\text{d}\tau \mathcal{K}_\mathcal{Q}(\tau)\mathcal{P}\hat{\rho}(t-\tau).
\end{equation}
So far we have only assumed a specific form for the projection operators and that the initial state is factorized and diagonal in the $\ket{\alpha}$ basis.
Next we will introduce a second standard assumption on the form of the system--environment interaction, typically made in the study of open quantum systems, to bring Eq.~(\ref{eq:generalized_master_equation_general}) to the simplified form that we have considered in the main text.

\subsection{Simplified decomposition of the total coherence--environment influence}
The form of the joint system--environment Hamiltonian is taken to be $\hat{H} = \hat{H}_S + \hat{H}_E + \hat{H}_I$, with $\hat{H}_S$ and $\hat{H}_E$ being the system and the environment Hamiltonians, respectively, and with $\hat{H}_I$ describing the interaction between the two. 
A standard assumption for open quantum systems is that $\hat{H}_I$ is given in a bilinear form
\begin{equation}
    \hat{H}_I = \sum_k \hat{S}_k\otimes\hat{B}_k,
\end{equation}
where $\hat{S}_k$ and $\hat{B}_k$ are system and environment operators, respectively. 
We furthermore assume that we have $\text{Tr}_E(\hat{B}_k\hat{\rho}_E) = 0$ for all $k$. 
A contribution stemming from a nonzero $\text{Tr}_E(\hat{B}_k\hat{\rho}_E)$ can always be absorbed into a redefined $\hat{H}_S$. 
Expanding the $\mathcal{PLP}\hat{\rho}(t)$ term directly yields now
\begin{equation}
    \mathcal{PLP}\hat{\rho}(t) = -i\sum_{\alpha,\beta}\braket{\alpha\vert\hat{\rho}_S(t)\vert\alpha}\braket{\beta\vert\text{Tr}_E([\hat{H}, \ketbra{\alpha}{\alpha}\otimes\hat{\rho}_E])\vert\beta}\ketbra{\beta}{\beta}\otimes\hat{\rho}_E,
\end{equation}
which can be shown to reduce to zero by noting that we have $\braket{\beta\vert\text{Tr}_E([\hat{H}, \ketbra{\alpha}{\alpha}\otimes\hat{\rho}_E])\vert\beta} = \braket{\beta\vert[\hat{H}_S, \ketbra{\alpha}{\alpha}]\vert\beta} = 0$ for all $\alpha$ and $\beta$. 
Equation~(\ref{eq:generalized_master_equation_general}) is therefore simplified further as
\begin{equation}\label{eq:memory_kernel_eom}
    \partial_t\mathcal{P}\hat{\rho}(t) = \int_0^t\text{d}\tau \mathcal{K}_\mathcal{Q}(\tau)\mathcal{P}\hat{\rho}(t-\tau).
\end{equation}
By tracing out the environment, we then find the generalized master equation governing the reduced system population dynamics:
\begin{equation}\label{eq:generalized_master_equation}
    \partial_t p_\beta(t) = \sum_{\alpha} \int_0^t\text{d}\tau \mathcal{K}^{(\mathcal{Q})}_{\alpha\to\beta}(\tau) p_\alpha(t - \tau),
\end{equation}
where we have set $p_\alpha(t) = \braket{\alpha\vert\hat{\rho}_S(t)\vert\alpha}$ and $\mathcal{K}^{(\mathcal{Q})}_{\alpha\to\beta}(t) = \braket{\beta\vert\text{Tr}_E(\mathcal{K}_\mathcal{Q}(t)[\ketbra{\alpha}{\alpha}\otimes\hat{\rho}_E])\vert\beta}$.

Now we note that
\begin{align}
    \partial_\tau\left(e^{\mathcal{QL}\tau}e^{\mathcal{RL}(t - \tau)}\right) &= e^{\mathcal{QL}\tau}\mathcal{QL}e^{\mathcal{RL}(t - \tau)} - e^{\mathcal{QL}\tau}\mathcal{RL}e^{\mathcal{RL}(t - \tau)} \\
    &= e^{\mathcal{QL}\tau}(\mathcal{Q} - \mathcal{R})\mathcal{L}e^{\mathcal{RL}(t - \tau)} \\
    &= e^{\mathcal{QL}\tau}\mathcal{CL}e^{\mathcal{RL}(t - \tau)},
\end{align}
which we then integrate both sides with respect to $\tau$ from 0 to $t$ to find
\begin{equation}
    e^{\mathcal{QL}t} = e^{\mathcal{RL}t} + \int_0^t\text{d}\tau e^{\mathcal{QL}\tau}\mathcal{CL}e^{\mathcal{RL}(t - \tau)}.
\end{equation}
Using this general operator identity allows us to expand $\mathcal{K}_\mathcal{Q}(t)$ as
\begin{equation}\label{eq:general_decomposition}
    \mathcal{K}_\mathcal{Q}(t) = \mathcal{K}_\mathcal{R}(t) + \Delta\mathcal{K}^{(\text{coh})}(t),
\end{equation}
where we have denoted 
\begin{equation}
    \mathcal{K}_\mathcal{R}(t) = \mathcal{PL}e^{\mathcal{RL}t}\mathcal{RL}
\end{equation}
and
\begin{equation}
    \Delta\mathcal{K}^{(\text{coh})}(t) = \mathcal{PL}e^{\mathcal{RL}t}\mathcal{CL} + \int_0^t\text{d}\tau \mathcal{PL}e^{\mathcal{QL}\tau}\mathcal{CL}e^{\mathcal{RL}(t - \tau)}\mathcal{(R+C)L}. 
\end{equation}
We have therefore decomposed the total influence into two terms: the environmental influence excluding system coherences $\mathcal{K}_\mathcal{R}(t)$, and the influence that is contingent on system coherences $\Delta\mathcal{K}^{(\text{coh})}(t)$. 
We therefore find that the memory kernel elements of the population master equation in Eq.\ (\ref{eq:generalized_master_equation}) may be expanded as
\begin{equation}\label{eq:memory_kernel_elements_expansion}
    \mathcal{K}^{(\mathcal{Q})}_{\alpha\to\beta}(t) = \mathcal{K}^{(\mathcal{R})}_{\alpha\to\beta}(t) + \Delta\mathcal{K}^{(\text{coh})}_{\alpha\to\beta}(t),
\end{equation}
where we have $\mathcal{K}^{(\mathcal{R})}_{\alpha\to\beta}(t) = \braket{\beta\vert\text{Tr}_E(\mathcal{K}_\mathcal{R}(t)[\ketbra{\alpha}{\alpha}\otimes\hat{\rho}_E])\vert\beta}$ and $\Delta\mathcal{K}^{(\text{coh})}_{\alpha\to\beta}(t) = \braket{\beta\vert\text{Tr}_E(\Delta\mathcal{K}^{(\text{coh})}(t)[\ketbra{\alpha}{\alpha}\otimes\hat{\rho}_E])\vert\beta}$. 
It should be noted that the expansion in Eq.\ (\ref{eq:memory_kernel_elements_expansion}) does not rely on the assumptions we have made on the initial state or the form of $\hat{H}$, and is the result of a general operator identity. 
This means that the expansion may be considered alongside the inhomogeneous term that stems from initial system--environment correlations or initial system coherences, and similarly for a nonzero $\mathcal{PLP}\hat{\rho}(t)$ term. 
We have imposed these assumptions to simplify the presentation of our analysis.

Finally, we now introduce the assumption that the basis states $\{\ket{\alpha}\}$ are the eigenstates of $\hat{H}_S$, which is what we have assumed in the main text. 
We may then note that
\begin{equation}
    \mathcal{CLP}\hat{\rho}(t) = -i\sum_{\alpha}\sum_{\beta\neq\gamma}\braket{\alpha\vert\hat{\rho}_S(t)\vert\alpha}\braket{\beta\vert\text{Tr}_E([\hat{H}, \ketbra{\alpha}{\alpha}\otimes\hat{\rho}_E])\vert\gamma}\ketbra{\beta}{\gamma}\otimes\hat{\rho}_E
\end{equation}
will reduce to zero, as we have $\text{Tr}_E([\hat{H}, \ketbra{\alpha}{\alpha}\otimes\hat{\rho}_E]) = [\hat{H}_S, \ketbra{\alpha}{\alpha}] = 0$ for all $\alpha$. 
The contribution of eigenstate coherences to population transfer therefore reduces to a single term:
\begin{equation}
    \Delta\mathcal{K}^{(\text{coh})}(t)\mathcal{P} = \int_0^t\text{d}\tau \mathcal{PL}e^{\mathcal{QL}\tau}\mathcal{CL}e^{\mathcal{RL}(t - \tau)}\mathcal{RLP},
\end{equation}
such that $\Delta\mathcal{K}^{(\text{coh})}_{\alpha\to\beta}(t) = \braket{\beta\vert\text{Tr}_E(\int_0^t\text{d}\tau \mathcal{PL}e^{\mathcal{QL}\tau}\mathcal{CL}e^{\mathcal{RL}(t - \tau)}\mathcal{RL}[\ketbra{\alpha}{\alpha}\otimes\hat{\rho}_E])\vert\beta}$.

\subsection{Energy transfer rates in the steady-state limit}
In the steady state where $t\to\infty$ and $\mathcal{L}\hat{\rho}_{ss} = 0$, Eq.~(\ref{eq:general_decomposition}) may be integrated to find
\begin{equation}\label{eq:steady_state_simplified_identity}
    K_\mathcal{Q} = K_\mathcal{R} + \Delta K^{(\text{coh})},
\end{equation}
where we have introduced $K_X = \int_0^\infty\text{d}\tau \mathcal{K}_X(\tau) = \lim_{\epsilon\to0^+}\mathcal{PL}\frac{-1}{X\mathcal{L} - \epsilon}X\mathcal{L}$ for $X = \mathcal{Q}, \mathcal{R}$, and $\Delta K^{(\text{coh})} = \int_0^\infty\text{d}t\Delta\mathcal{K}^{(\text{coh})}(t)$. 
The difference $K_\mathcal{Q} - K_\mathcal{R}$ will therefore be nonzero only when environment-induced coherence impacts energy transfer in the stationary state. 
We furthermore have
\begin{align}
    (K_\mathcal{Q} - K_\mathcal{R})\mathcal{P} &= \int_0^\infty\text{d}\tau \int_0^\tau\text{d}s \mathcal{PL}e^{\mathcal{QL}s}\mathcal{CL}e^{\mathcal{RL}(\tau-s)}\mathcal{RLP} \\
    &= \int_0^\infty\text{d}s \int_s^\infty\text{d}\tau \mathcal{PL}e^{\mathcal{QL}s}\mathcal{CL}e^{\mathcal{RL}(\tau-s)}\mathcal{RLP} \\
    &= \int_0^\infty\text{d}s \int_0^\infty\text{d}u \mathcal{PL}e^{\mathcal{QL}s}\mathcal{CL}e^{\mathcal{RL}u}\mathcal{RLP} \\
    &= \lim_{\epsilon,\epsilon^\prime\to0^+}\mathcal{PL}\frac{-1}{\mathcal{QL}-\epsilon}\mathcal{CL}\frac{-1}{\mathcal{RL}-\epsilon^\prime}\mathcal{RLP}.
\end{align}
Generalized population transfer rates in the reduced system may therefore be calculated as
\begin{equation}
    k_{\alpha\to\beta}^{(X)} = \braket{\beta\vert\text{Tr}_E(K_X[\ketbra{\alpha}{\alpha}\otimes\hat{\rho}_E])\vert\beta},
\end{equation}
for $\alpha\neq\beta$ and $X = \mathcal{Q},\mathcal{R}$, and their difference must stem from environment-induced coherence.

\section{Open quantum system dynamics for a dimer model}
The Frenkel exciton Hamiltonian for the electronically coupled dimer is given as
\begin{equation}
    \hat{H}_S = \frac{\Delta\epsilon}{2}(\ketbra{e_1}{e_1} - \ketbra{e_2}{e_2}) + V(\ketbra{e_1}{e_2} + \ketbra{e_2}{e_1}),
\end{equation}
which we may rewrite as $\hat{H}_S = \frac{\Delta\epsilon}{2}(\hat{n}_1 - \hat{n}_2) + V(\hat{\sigma}_1^\dagger\hat{\sigma}_2 + \hat{\sigma}_1\hat{\sigma}_2^\dagger)$,
with $\hat{n}_i = \hat{\sigma}_i^\dagger\hat{\sigma}_i$ as the site population operator for sites $i=1,2$ and $\hat{\sigma}_i$ ($\hat{\sigma}_i^\dagger$) is the site excitation annihilation (creation) operator satisfying $\hat{\sigma}_i\ket{e_i} = \ket{0}$ ($\hat{\sigma}_i^\dagger\ket{0} = \ket{e_i}$), where $\ket{0}$ is the zero-excitation state. 
The total Hamiltonian $\hat{H}$ of the dimer coupled to the displacements of the normal modes of a thermal environment is then taken to be of the form (setting $\hbar = 1$)
\begin{equation}\label{eq:total_dimer_hamiltonian}
    \hat{H} = \hat{H}_S + \sum_{i,k}\left[\omega_{k}\hat{b}_{i,k}^\dagger\hat{b}_{i,k} + g_{k}\hat{n}_{i}(\hat{b}_{i,k}^\dagger + \hat{b}_{i,k})\right],
\end{equation}
where $\hat{b}_{i,k}^\dagger$ ($\hat{b}_{i,k}$) is the bosonic creation (annihilation) operator satisfying the commutation relation $[\hat{b}_{i,k},\hat{b}_{j,k^\prime}^\dagger] = \delta_{i,j}\delta_{k,k^\prime}$. 
We have assumed the local vibrational environments to be identical such that we can make a unitary change of basis into global modes by setting $\hat{b}_{\pm,k} = (\hat{b}_{1,k}\pm\hat{b}_{2,k})/\sqrt{2}$ so that we may rewrite $\hat{H}$ as
\begin{equation}
    \hat{H} = \hat{H}_S + \sum_{k}\left\{\omega_{k}(\hat{b}_{+,k}^\dagger\hat{b}_{+,k} + \hat{b}_{-,k}^\dagger\hat{b}_{-,k}) + \frac{g_k}{\sqrt{2}}\left[(\hat{n}_1 + \hat{n}_2)(\hat{b}_{+,k}^\dagger + \hat{b}_{+,k}) + (\hat{n}_1 - \hat{n}_2)(\hat{b}_{-,k}^\dagger + \hat{b}_{-,k})\right]\right\}.
\end{equation}

We have restricted our analysis to the single-excitation manifold, where the coupling operator $\hat{n}_1 + \hat{n}_2$ reduces to the identity, such that the corresponding interaction term results in a simple energy shift. 
The single-excitation manifold dynamics are therefore only influenced by the relative displacement modes, such that we may consider the reduced description
\begin{equation}
    \hat{H} = \hat{H}_S + \sum_{k}\left[\omega_{k}\hat{b}_{k}^\dagger\hat{b}_{k} + \frac{g_k}{\sqrt{2}}(\hat{n}_1 - \hat{n}_2)(\hat{b}_{k}^\dagger + \hat{b}_{k})\right],
\end{equation}
where we have dropped the $\pm$ index as it is no longer required. 
We finally arrive at the form
\begin{equation}\label{eq:reduced_hamiltonian}
    \hat{H} = \frac{\Delta\epsilon}{2}\hat{\sigma}_z + V\hat{\sigma}_x + \sum_{k}\left[\omega_{k}\hat{b}_{k}^\dagger\hat{b}_{k} + \frac{g_k}{\sqrt{2}}\hat{\sigma}_z(\hat{b}_{k}^\dagger + \hat{b}_{k})\right],
\end{equation}
where $\hat{\sigma}_z = \hat{n}_1 - \hat{n}_2$ and $\hat{\sigma}_x = \hat{\sigma}_1^\dagger\hat{\sigma}_2 + \hat{\sigma}_1\hat{\sigma}_2^\dagger$.
We can therefore simplify our model by reducing the number of environments from two to one without making any approximations, and so Eq.~(\ref{eq:reduced_hamiltonian}) is the form we have implemented numerically.

We define interaction picture operators with respect to $\hat{H}_0 = \hat{H}_S + \sum_{k}\omega_{k}\hat{b}_{k}^\dagger\hat{b}_{k}$ such that we have $\tilde{O}(t) = e^{i\hat{H}_0t}\hat{O}e^{-i\hat{H}_0t}$, where $\hat{O}$ is a Schrödinger picture operator and the tilde indicates an operator in the interaction picture. 
The dynamics of the density matrix in the interaction picture is now determined by the Liouville–von Neumann equation of motion
\begin{align}
    \partial_t\tilde{\rho}(t) &= -i\sum_{k}\frac{g_k}{\sqrt{2}}[\tilde{\sigma}_z(t)\otimes(\tilde{b}_{k}^\dagger(t) + \tilde{b}_{k}(t)),\tilde{\rho}(t)] \\
    &= \mathcal{L}(t)\tilde{\rho}(t).
\end{align}
For initial conditions $\tilde{\rho}(0)$ at time $t=0$, the formal solution is
\begin{equation}
    \tilde{\rho}(t) = \mathcal{T}\{e^{\int_0^t\text{d}\tau\mathcal{L}(\tau)}\}\tilde{\rho}(0),
\end{equation}
where $\mathcal{T}$ indicates time ordering, such that the reduced system dynamics are obtained by the partial trace over the environmental normal modes:
\begin{equation}
    \tilde{\rho}_S(t) = \text{Tr}_E(\mathcal{T}\{e^{\int_0^t\text{d}\tau\mathcal{L}(\tau)}\}\tilde{\rho}(0)).
\end{equation}

Assuming factorized initial conditions $\tilde{\rho}(0) = \tilde{\rho}_S(0)\otimes\tilde{\rho}_E(0)$, where we have $\hat{\rho}_E = e^{-\beta\hat{H}_E} / \text{Tr}(e^{-\beta\hat{H}_E})$ with $\beta = (k_\text{B}T)^{-1}$ and $\hat{H}_E = \sum_{k}\omega_{k}\hat{b}_{k}^\dagger\hat{b}_{k}$, it can be shown that the reduced system dynamics may be rewritten in terms of a Feynman–Vernon influence functional
\begin{equation}\label{eq:f_v_reduced_dynamics}
    \tilde{\rho}_S(t) = \mathcal{T}\{e^{\mathcal{F}(t)}\}\tilde{\rho}_S(0),
\end{equation}
where we have introduced the superoperator
\begin{equation}\label{eq:f_v_functional}
    \mathcal{F}(t) = -\frac{1}{2}\int_0^t\text{d}t_1\tilde{\sigma}_z^\times(t_1)\int_0^{t_1}\text{d}t_2 \left[C(t_1-t_2)\tilde{\sigma}_z^\text{L}(t_2) - C^*(t_1-t_2)\tilde{\sigma}_z^\text{R}(t_2)\right],
\end{equation}
with $C(t) = \frac{1}{\pi}\int_0^\infty\text{d}\omega J(\omega)[(1 + \frac{2}{e^{\beta\omega} - 1})\cos(\omega t) - i\sin(\omega t)]$ being the environment correlation function, where $J(\omega)$ is the spectral density, and $\hat{\sigma}_z^\times\hat{\rho} = [\hat{\sigma}_z, \hat{\rho}]$, $\hat{\sigma}_z^\text{L}\hat{\rho} = \hat{\sigma}_z\hat{\rho}$, $\hat{\sigma}_z^\text{R}\hat{\rho} = \hat{\rho}\hat{\sigma}_z$.

\subsection{No environment-induced coherence in the case of zero detuning}
In the case of no detuning ($\Delta\epsilon = 0$), we can show that the influence of the environment does not generate excitonic coherence when the initial state is diagonal in the eigenstate basis.
This can be demonstrated by using the geometric symmetry that exists, whereby exchanging the coordinates of the sites leaves $\hat{H}$ invariant.
According to Wigner's theorem, there exists a unitary operator $\hat{P} = \hat{P}_S\otimes\hat{P}_E$, with $\hat{P}_S$ and $\hat{P}_E$ acting on the system and environment respectively, that represents the aforementioned symmetry transformation such that $\hat{P}\hat{H}\hat{P}^{-1} = \hat{H}$.
Specifically, the transformation rules are: $\hat{P}_S\hat{\sigma}_{1}\hat{P}^{-1}_S = \hat{\sigma}_{2}$, $\hat{P}_S\hat{\sigma}_{2}\hat{P}^{-1}_S = \hat{\sigma}_{1}$, $\hat{P}_E\hat{b}_{1,k}\hat{P}^{-1}_E = \hat{b}_{2,k}$, $\hat{P}_E\hat{b}_{2,k}\hat{P}^{-1}_E = \hat{b}_{1,k}$. 
Note that the local environments must be identical for this symmetry to hold. 
The eigenstates of $\hat{H}_S$ therefore have an even or odd parity with respect to $\hat{P}_S$.
We have
\begin{align}
    \hat{P}_S\ket{1} &= \frac{1}{\sqrt{2}}(\hat{P}_S\ket{e_1} - \hat{P}_S\ket{e_2}) = -\ket{1}, \\
    \hat{P}_S\ket{2} &= \frac{1}{\sqrt{2}}(\hat{P}_S\ket{e_1} + \hat{P}_S\ket{e_2}) = +\ket{2}.
\end{align}
Given the initial condition $\hat{\rho}(0) = (p_1\ketbra{1}{1}+p_2\ketbra{2}{2})\otimes\hat{\rho}_E$ the state at time $t\geq0$ is given by $\hat{\rho}(t) = e^{-i\hat{H}t}\hat{\rho}(0)e^{i\hat{H}t}$.
The initial condition is invariant with respect to $\hat{P}$ and the dynamics preserve this property for all times:
\begin{align}
    \hat{P}\hat{\rho}(t)\hat{P}^{-1} &= \hat{P}e^{-i\hat{H}t}\hat{P}^{-1}\hat{P}\hat{\rho}(0)\hat{P}^{-1}\hat{P}e^{i\hat{H}t}\hat{P}^{-1} \\
    &= e^{-i\hat{H}t}\left[\hat{P}_S(p_1\ketbra{1}{1}+p_2\ketbra{2}{2})\hat{P}^{-1}_S\otimes\hat{P}_E\hat{\rho}_E\hat{P}^{-1}_E\right]e^{i\hat{H}t} \\
    &= \hat{\rho}(t).
\end{align}
Now we consider the partial trace over the environment, $\hat{\rho}_S(t) = \text{Tr}_E(\hat{\rho}(t))$, which obeys the symmetry relation
\begin{align}
    \hat{\rho}_S(t) &= \text{Tr}_E(\hat{P}\hat{\rho}(t)\hat{P}^{-1}) \\
    &= \hat{P}_S\text{Tr}_E((\hat{I}_S\otimes\hat{P}_E)\hat{\rho}(t)(\hat{I}_S\otimes\hat{P}^{-1}_E))\hat{P}^{-1}_S \\
    &= \hat{P}_S\text{Tr}_E(\hat{\rho}(t))\hat{P}^{-1}_S \\
    &= \hat{P}_S\hat{\rho}_S(t)\hat{P}^{-1}_S,
\end{align}
where we have used the fact that the partial trace is invariant under cyclic permutations with respect to the operator $\hat{I}_S\otimes\hat{P}_E$. 
Finally, we expand the reduced state in the eigenstate basis as
\begin{equation}
    \hat{\rho}_S(t) = \rho_{11}(t)\ketbra{1}{1} + \rho_{12}(t)\ketbra{1}{2} + \rho_{21}(t)\ketbra{2}{1} + \rho_{22}(t)\ketbra{2}{2},
\end{equation}
with $\rho_{ij}(t) = \braket{i\vert\hat{\rho}_S(t)\vert j}$.
We therefore see that the exchange symmetry implies $\rho_{12}(t) = \braket{1\vert\hat{P}_S\hat{\rho}_S(t)\hat{P}^{-1}_S\vert 2} = -\rho_{12}(t)$, which can only be satisfied if $\rho_{12}(t) = 0$ for all $t$. 
Therefore, given the initial condition $\hat{\rho}(0) = (p_1\ketbra{1}{1}+p_2\ketbra{2}{2})\otimes\hat{\rho}_E$, we find that the ensuing dynamics cannot generate any coherence in the eigenstate basis. 
The physical manifestation of the dynamic constraints imposed by the symmetry is therefore a destructive interference suppressing generation of excitonic coherence.

\section{Hierarchical equations of motion}\label{sec:heom}
The hierarchical equations of motion (HEOM) may be derived by assuming that the environment correlation function $C(t)$, defined immediately after Eq.~(\ref{eq:f_v_functional}), is of the form of an exponential sum:
\begin{equation}\label{eq:correlation_exponential_expansion}
    C(t) = \sum_{k=0}^K c_{k} e^{-\nu_{k}t} + \sum_{k^\prime=0}^{K^\prime}\left(c_{k^\prime,+} e^{-\nu_{k^\prime,+} t} + c_{k^\prime,-} e^{-\nu_{k^\prime,-} t}\right),
\end{equation}
where $(\nu_{k^\prime,+})^* = \nu_{k^\prime,-}$. 
The number of terms with real-valued $\nu_k$ is $K+1$ and the number of terms with complex-valued $\nu_{k^\prime,\pm}$ is $2(K^\prime+1)$. 
The coefficients of this expansion for the overdamped and the underdamped Brownian oscillator spectral densities considered in the main text are given in Sec.\ \ref{sec:heom_coeffs}. 
We now introduce superoperators
\begin{gather}
    \hat{\Theta}_{k} = c_{k}\hat{\sigma}_z^\text{L} - (c_{k})^*\hat{\sigma}_z^\text{R}, \\
    \hat{\Theta}_{k,\pm} = c_{k,\pm}\hat{\sigma}_z^\text{L} - (c_{k,\mp})^*\hat{\sigma}_z^\text{R},
\end{gather}
such that the HEOM in the Schrödinger picture is
\begin{equation}\label{eq:heom_form}
    \partial_t\hat{\rho}_{\vec{n}} = (-i\hat{H}_S^\times - \nu_{\vec{n}})\hat{\rho}_{\vec{n}} + \frac{1}{\sqrt{2}}\sum_{r\in\mathcal{S}}(\hat{\sigma}_z^\times\hat{\rho}_{\vec{n}+1_r} - n_r\hat{\Theta}_r\hat{\rho}_{\vec{n}-1_r}),
\end{equation}
where $\vec{n}$ is a multi-index that consists of non-negative integer-valued entries, $n_r$, and denotes the position of the auxiliary $\hat{\rho}_{\vec{n}}$ in the hierarchy, and $\nu_{\vec{n}} = \sum_{r\in\mathcal{S}} n_r\nu_r$. 
The multi-index that exclusively consists of zero entries corresponds to the reduced system density matrix.
The set $\mathcal{S}$ is composed of elements corresponding to the labels of the terms in the expansion in Eq.~(\ref{eq:correlation_exponential_expansion}), e.g., $r = (k,\pm)$, and $\vec{n}\pm1_r$ denotes a multi-index with unity added or subtracted in the $r$-th entry of $\vec{n}$. 
The hierarchy is formally infinite, so a truncation is introduced in its numerical implementation by setting auxiliary operators beyond a preset threshold to zero \cite{Ishizaki2005Dec}. 
Finally, we utilize the scaling proposed in Ref.~\cite{Shi2009Feb} to improve the numerical precision of our implementation. 
The scaling is achieved by sending $\hat{\rho}_{\vec{n}} \to \frac{1}{\prod_{r\in\mathcal{S}}\sqrt{n_r!s_r^{n_r}}}\hat{\rho}_{\vec{n}}$, where we have
\begin{gather}
    s_k = \vert c_k\vert, \\
    s_{k,\pm} = \text{max}(\vert c_{k,+}\vert, \vert c_{k,-}\vert),
\end{gather}
so that the rescaled HEOM will be
\begin{equation}\label{eq:rescaled_heom_form}
    \partial_t\hat{\rho}_{\vec{n}} = (-i\hat{H}_S^\times - \nu_{\vec{n}})\hat{\rho}_{\vec{n}} + \frac{1}{\sqrt{2}}\sum_{r\in\mathcal{S}}(\sqrt{(n_r + 1)s_r}\hat{\sigma}_z^\times\hat{\rho}_{\vec{n}+1_r} - \sqrt{\frac{n_r}{s_r}}\hat{\Theta}_r\hat{\rho}_{\vec{n}-1_r}).
\end{equation}

\subsection{Environment correlation expansion coefficients}\label{sec:heom_coeffs}
For a Drude–Lorentz spectral density
\begin{equation}
    J(\omega) = 2\lambda\gamma\frac{\omega}{\omega^2 + \gamma^2},
\end{equation}
we have used the Matsubara expansion coefficients in our HEOM.
They are:
\begin{align}
    \nu_{0} &= \gamma, \\
    c_{0} &= \lambda\gamma\left(\cot\left(\frac{\beta\gamma}{2}\right) - i\right), \\
    \nu_{k} &= \frac{2\pi k}{\beta}, \\
    c_{k} &= \frac{4\lambda\gamma}{\beta}\frac{\nu_{k}}{\nu_{k}^2 - \gamma^2}, 
\end{align}
for $k=1,2,3,\ldots$.

For an underdamped Brownian oscillator spectral density
\begin{equation}
    J(\omega) = 2\lambda\gamma\Omega^2\frac{\omega}{(\Omega^2 - \omega^2)^2 + \gamma^2\omega^2},
\end{equation}
we have used the Matsubara expansion coefficients in our HEOM.
They are:
\begin{align}
    \nu_{\pm} &= \frac{\gamma}{2} \pm i\omega^\prime, \\
    c_{\pm} &= \pm\frac{\lambda\Omega^2}{2\omega^\prime}\left(1 + i\cot\left(\frac{\beta\nu_{\pm}}{2}\right)\right), \\
    \nu_{k} &= \frac{2\pi k}{\beta}, \\
    c_{k} &= -\frac{4\lambda\gamma\Omega^2}{\beta}\frac{\nu_{k}}{(\nu_{+}^2-\nu_{k}^2)(\nu_{-}^2-\nu_{k}^2)}, 
\end{align}
where $\omega^\prime = \sqrt{\Omega^2 - \frac{\gamma^2}{4}}$ and $k=1,2,3,\ldots$.

\section{Memory kernel formulation of HEOM}
The formalism we have put forth is general for any open quantum system dynamics and may therefore be used in conjunction with any non-perturbative approach to computing the reduced system dynamics. 
In this section we demonstrate how the Nakajima--Zwanzig memory kernel formulation can be mapped to the HEOM formulation as explained in Ref.~\cite{Zhang2016May}. 
Using projection operators we can write an equation of motion for the reduced system density matrix:
\begin{align}
    \partial_t\hat{\rho}_S(t) &= \partial_t\text{Tr}_E(\hat{\rho}(t)) \\
    &= \text{Tr}_E(\mathcal{L}\hat{\rho}(t)) \\
    &= \text{Tr}_E(\mathcal{L}(\mathcal{P} + \mathcal{C} + \mathcal{R})\hat{\rho}(t)) \\
    &= -i\hat{H}_S^\times\hat{\rho}_S(t) + \text{Tr}_E(\mathcal{LR}\hat{\rho}(t)).
\end{align}
Recalling now the definition of the Feynman--Vernon formulation of the reduced system dynamics given in Eq.~(\ref{eq:f_v_reduced_dynamics}) and writing it in the Schrödinger picture as $\hat{\rho}_S(t) = e^{-i\hat{H}_S^\times t}\mathcal{T}\left\{e^{\mathcal{F}(t)}\right\}\tilde{\rho}_S(0)$, we also will have the equation of motion
\begin{equation}
    \partial_t\hat{\rho}_S(t) = -i\hat{H}_S^\times\hat{\rho}_S(t) + e^{-i\hat{H}_S^\times t}\mathcal{T}\left\{\partial_t e^{\mathcal{F}(t)}\right\}\tilde{\rho}_S(0),
\end{equation}
such that we have
\begin{equation}
    \text{Tr}_E(\mathcal{LR}\hat{\rho}(t)) = e^{-i\hat{H}_S^\times t}\mathcal{T}\left\{\partial_t e^{\mathcal{F}(t)}\right\}\tilde{\rho}_S(0).
\end{equation}
The right-hand side may now be expanded in terms of the standard HEOM auxiliary density operators to arrive at the hierarchy presented in Sec. \ref{sec:heom}.

We first introduce the vector $\vec{\rho}(t)$ whose elements consist of the reduced density matrix $\hat{\rho}_S(t)$ as its first entry, followed by all ensuing auxiliary density operators $\hat{\rho}_{\vec{n}}(t)$ according to an ordering that is specified by the numerical implementation. 
The hierarchy determined by the recursion in Eq.~(\ref{eq:rescaled_heom_form}) is now represented by the linear map $L$ such that we have the equation of motion
\begin{equation}
    \partial_t\vec{\rho}(t) = L\vec{\rho}(t),
\end{equation}
with the initial condition $\vec{\rho}(0)$ determined by $\hat{\rho}_S(0)$ and all auxiliary density operators initially set to zero. 
Recalling now the definition of the projection operators we have used in the main text, we have $\mathcal{P}\hat{\rho}(t) = \sum_\alpha\braket{\alpha\vert\hat{\rho}_S(t)\vert\alpha}\ketbra{\alpha}{\alpha}\otimes\hat{\rho}_E$ (with $\ket{\alpha}$ being the reduced system eigenstates), where a partial trace over the environment yields
\begin{equation}
    \text{Tr}_E(\mathcal{P}\hat{\rho}(t)) = \sum_\alpha\braket{\alpha\vert\hat{\rho}_S(t)\vert\alpha}\ketbra{\alpha}{\alpha},
\end{equation}
implying the existence of the corresponding map $P$ that acts on $\vec{\rho}(t)$ as
\begin{equation}
    P\vec{\rho}(t) = 
    \begin{bmatrix}
        \sum_\alpha\braket{\alpha\vert\hat{\rho}_S(t)\vert\alpha}\ketbra{\alpha}{\alpha} \\ 0 \\ 0 \\ \vdots
    \end{bmatrix},
\end{equation}
such that it sets all auxiliary elements to zero. 
For the $\mathcal{C}$ projection we similarly have the corresponding $C$ map with
\begin{equation}
    C\vec{\rho}(t) = 
    \begin{bmatrix}
        \sum_{\alpha\neq\beta}\braket{\alpha\vert\hat{\rho}_S(t)\vert\beta}\ketbra{\alpha}{\beta} \\ 0 \\ 0 \\ \vdots
    \end{bmatrix}.
\end{equation}
We therefore may define the complementary map $Q = I - P$, with $I$ being the identity map, and $R = Q - C$. 
We then have the corresponding coupled equations of motion:
\begin{align}
    \partial_tP\vec{\rho}(t) &= PLP\vec{\rho}(t) + PLQ\vec{\rho}(t), \label{eq:heom_naka_zwan_1} \\
    \partial_tQ\vec{\rho}(t) &= QLP\vec{\rho}(t) + QLQ\vec{\rho}(t), \label{eq:heom_naka_zwan_2}
\end{align}
where the formal solution for $Q\vec{\rho}(t)$ is
\begin{equation}
    Q\vec{\rho}(t) = e^{QLt}Q\vec{\rho}(0) + \int_0^t\text{d}\tau e^{QL\tau}QLP\vec{\rho}(t-\tau).
\end{equation}
The generalized master equation for $P\vec{\rho}(t)$ is now
\begin{equation}\label{eq:formal_eom_for_the_prho}
    \partial_tP\vec{\rho}(t) = PLe^{QLt}Q\vec{\rho}(0) + \int_0^t\text{d}\tau \mathcal{K}_Q(\tau)P\vec{\rho}(t-\tau),
\end{equation}
where we have used the fact that $PLP = 0$, and we denote the total memory kernel by $\mathcal{K}_Q(\tau) = PLe^{QL\tau}QL$.

The form of Eq.~(\ref{eq:formal_eom_for_the_prho}) neatly separates contributions to the energy transfer dynamics that stem from any potential initial coherences and those due to environment-induced coherence. 
This has potential utility for the study of spectroscopic pump--probe experiments on light-harvesting antennae, which rely on coherent laser light for the initial photoexcitation, and may therefore introduce excitonic coherence as an unwanted experimental artifact.

We have assumed in the main text that there are no initial coherences and taken the steady-state limit, such that the right-hand side of Eq.~(\ref{eq:formal_eom_for_the_prho}) reduces to
\begin{equation}
    \int_0^\infty\text{d}\tau \mathcal{K}_Q(\tau)P\vec{\rho}_{ss} = \sum_{\alpha,\beta} \braket{\beta\vert\int_0^\infty\text{d}\tau \mathcal{K}_Q(\tau)[\ketbra{\alpha}{\alpha}]\vert\beta}\braket{\alpha\vert P\vec{\rho}_{ss}\vert\alpha}\ketbra{\beta}{\beta},
\end{equation}
where $L\vec{\rho}_{ss} = 0$. 
The rate of population transfer from state $\ket{\alpha}$ to state $\ket{\beta}$ in the steady state can now be calculated numerically as
\begin{equation}
    \braket{\beta\vert\int_0^\infty\text{d}\tau \mathcal{K}_Q(\tau)[\ketbra{\alpha}{\alpha}]\vert\beta} = \lim_{\epsilon\to 0^+}\braket{\beta\vert PL\frac{-1}{QL - \epsilon}QL[\ketbra{\alpha}{\alpha}]\vert\beta},
\end{equation}
which is what we solve numerically, as it reduces to solving a linear equation involving a sparse matrix representation of $QL$.
The environment relaxation rate is calculated similarly as $\braket{\beta\vert\int_0^\infty\text{d}\tau \mathcal{K}_R(\tau)[\ketbra{\alpha}{\alpha}]\vert\beta} = \lim_{\epsilon\to 0^+}\braket{\beta\vert PL\frac{-1}{RL - \epsilon}RL[\ketbra{\alpha}{\alpha}]\vert\beta}$.
A small positive $\epsilon$ value is introduced in our numerical implementation to guarantee analyticity of the reciprocal operator, as $QL$ and $RL$ are formally singular.

\begin{figure}
    \centering
    \includegraphics[width=\textwidth]{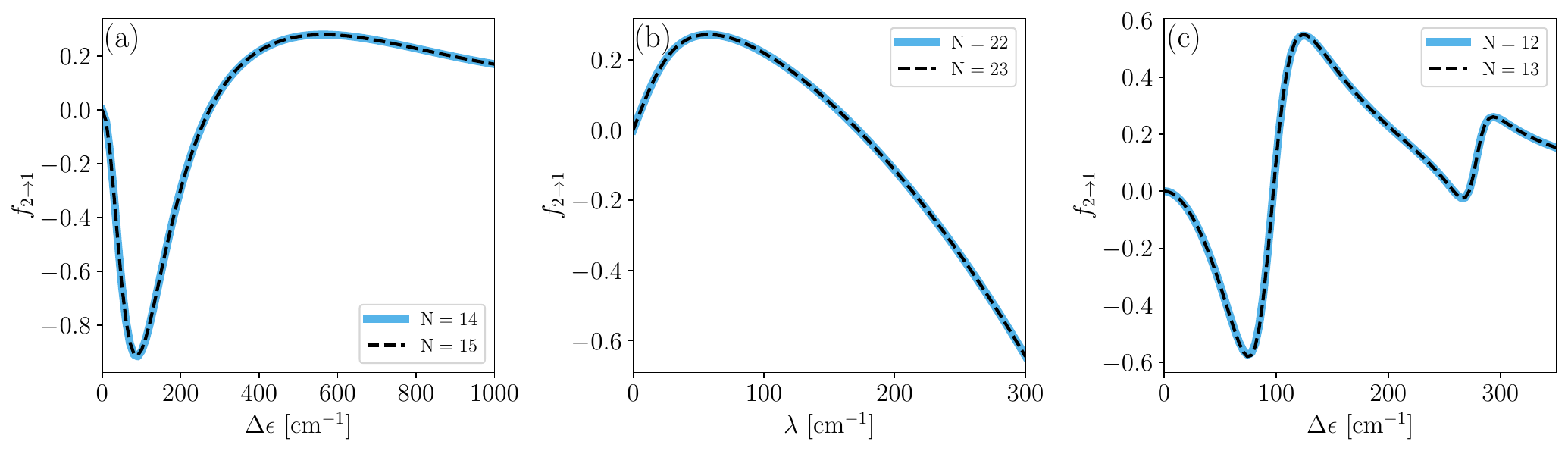}
    \caption{Here we demonstrate the convergence of the results presented in the main text by plotting $f_{2\to1}$ for two successive truncations of the HEOM, where $N$ is the truncation depth. 
    Here we have used $M = 1$. 
    (a) corresponds to Fig.~\ref{fig:rates}, (b) corresponds to Fig.~\ref{fig:rates_v_reorg}, and (c) corresponds to Fig.~\ref{fig:rates_with_ubo}. 
    A higher truncation depth is used for (b) due to the larger values of the reorganization energy $\lambda$. }
    \label{fig:convergence_figure}
\end{figure}

Figure~\ref{fig:convergence_figure} demonstrates the convergence of the $f_{2\to1}$ results discussed in the main text.
We found that the HEOM results presented in the main text converged by including one Matsubara mode ($M = 1$) in the hierarchy, and by setting $\epsilon = 10^{-5}$~cm$^{-1}$. 
We have accounted for neglected Matsubara modes in the Markovian limit, similar to the high-temperature approximation for HEOM derived in Ref.~\cite{Ishizaki2005Dec}.
We therefore will have the approximation
\begin{equation}
    \int_0^\infty\text{d}\tau \mathcal{K}_Q(\tau)P \approx P\Xi P + \lim_{\epsilon\to 0^+}P(L+\Xi)\frac{-1}{Q(L+\Xi) - \epsilon}Q(L+\Xi)P,
\end{equation}
where we have set $\Xi = -\frac{C_M}{2}\hat{\sigma}^\times_z \hat{\sigma}^\times_z$ with $C_M = \frac{2\lambda}{\beta\gamma} - i\lambda - \frac{c_0}{\nu_0} - \sum_{k=1}^M \frac{c_k}{\nu_k}$ in the overdamped case, and $C_M = \frac{2\lambda\gamma}{\beta\Omega^2} - i\lambda - \frac{c_+}{\nu_+} - \frac{c_-}{\nu_-} - \sum_{k=1}^M \frac{c_k}{\nu_k}$ in the underdamped case. 
This can be seen by setting $L\to L+\Xi$ in Eqs.~(\ref{eq:heom_naka_zwan_1}--\ref{eq:heom_naka_zwan_2}). 
We similarly have $\int_0^\infty\text{d}\tau \mathcal{K}_R(\tau)P \approx P\Xi P + \lim_{\epsilon\to 0^+}P(L+\Xi)\frac{-1}{R(L+\Xi) - \epsilon}R(L+\Xi)P$. 
Note that the $\Xi$ correction stems from the environment influence, $R$, so the fact that $P\Xi P \neq 0$ is not at odds with $PLP = PLC = CLP = 0$.

\clearpage
\section{Supplementary numerical results}

\subsection{Downhill excitonic energy transfer rates for several values of electronic coupling}
\begin{figure}[!htbp]
    \centering
    \includegraphics[width=0.87\linewidth]{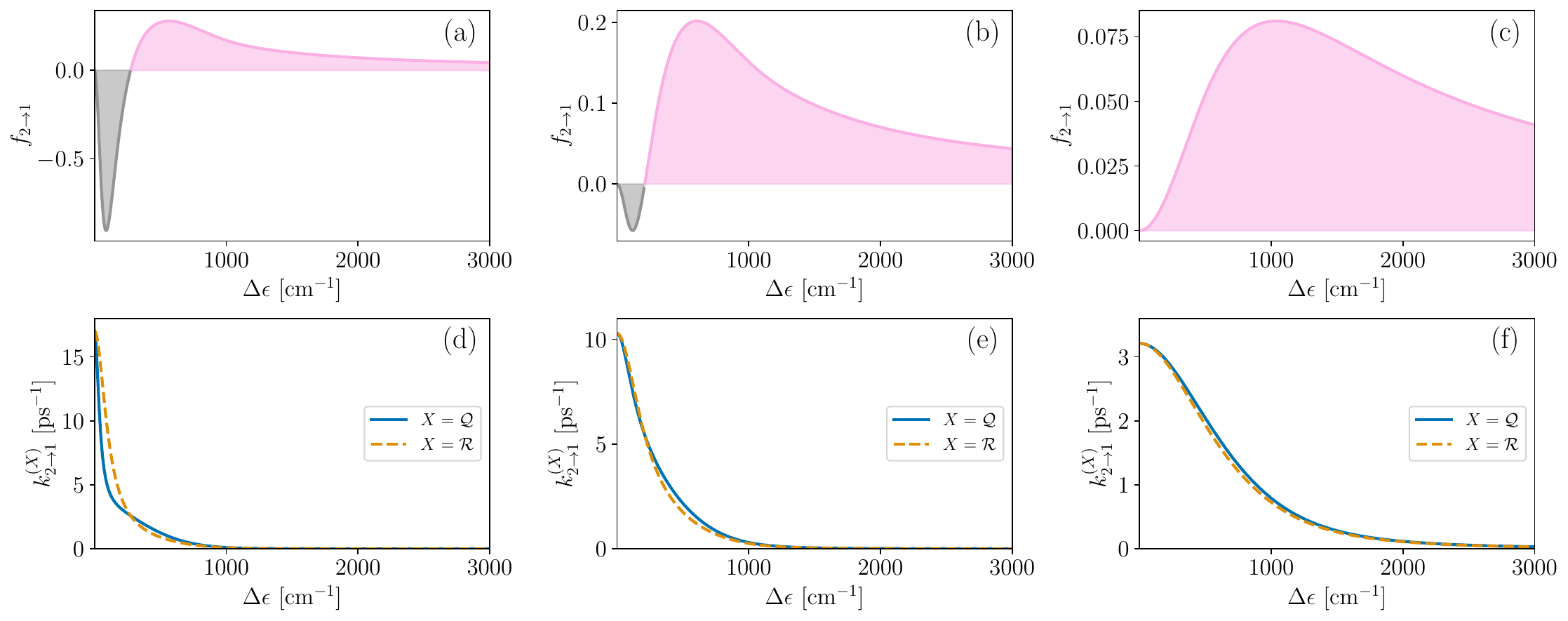}
    \caption{The figure of merit $f_{2\to1}$ (a–c) and the transfer rates (d–f) as a function of detuning in a dimer coupled to an overdamped Brownian oscillator, where we have set $V = 50$~cm$^{-1}$ (a), (d), $V = 100$~cm$^{-1}$ (b), (e), $V = 300$~cm$^{-1}$ (c), (f). 
    Truncation depth $N = 15$ used for all.
    The gray and pink regions correspond to regimes where the functional role of quantum coherence is to suppress and enhance the energy transfer process, respectively. 
    }
    \label{fig:figure_1_for_sm}
\end{figure}

Figure~\ref{fig:figure_1_for_sm} (similar to Fig.~\ref{fig:rates} in the main text) shows our numerical results for three different values of the electronic coupling $V$.

\begin{figure}[!htbp]
    \centering
    \includegraphics[width=0.87\linewidth]{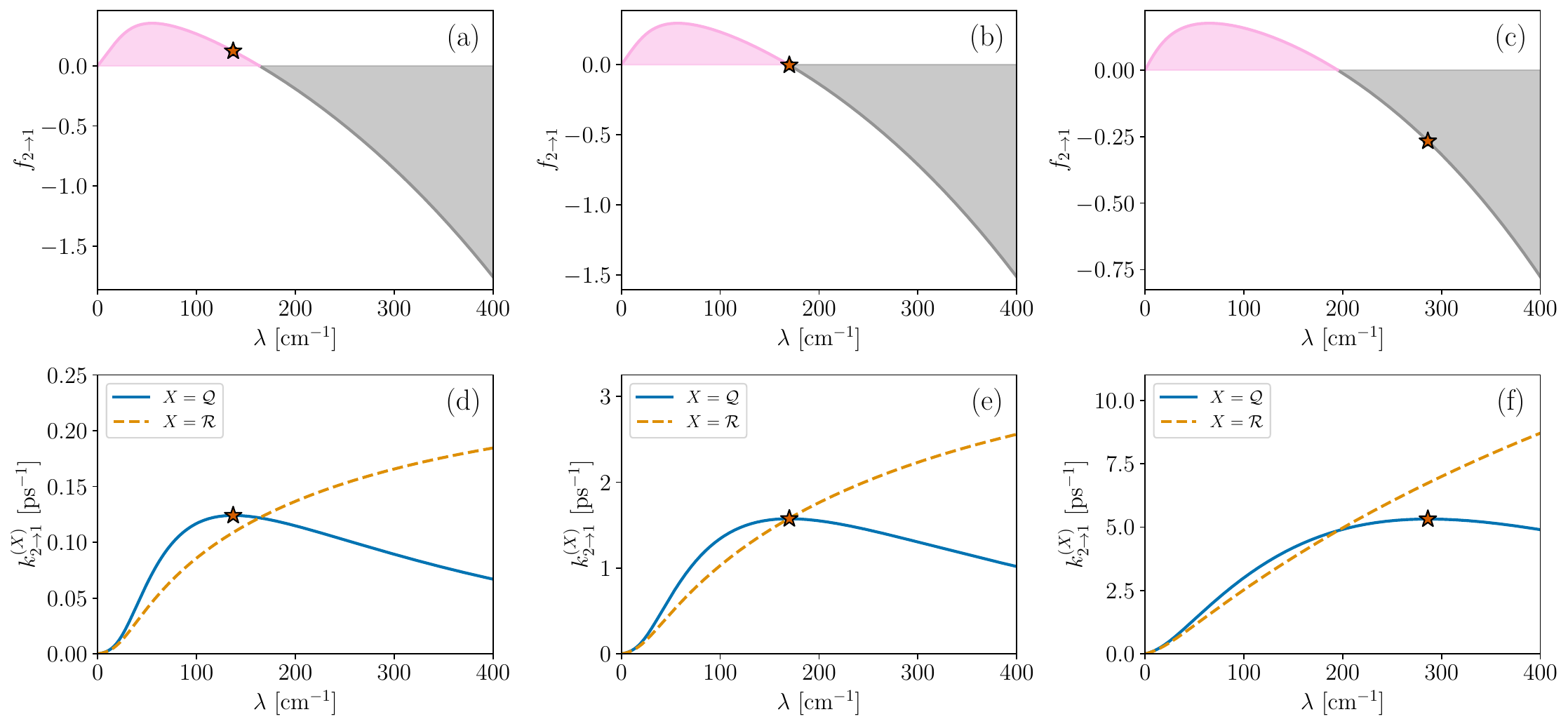}
    \caption{The figure of merit $f_{2\to1}$ (a–c) and the transfer rates (d–f) as a function of the reorganization energy of the overdamped Brownian oscillator, where we have fixed $\Delta\epsilon = 400$~cm$^{-1}$ and $V = 10$~cm$^{-1}$ (a), (d), $V = 40$~cm$^{-1}$ (b), (e), $V = 100$~cm$^{-1}$ (c), (f). 
    Truncation depth $N = 23$ used in all calculations. 
    The star marker indicates the point where the optimal value of $k^{(\mathcal{Q})}_{2\to1}$ is reached.
    This shows that the functional role of quantum coherence can be either enhancing, suppressing, or neither, when the system achieves its optimal total transfer rate. 
    }
    \label{fig:figure_3_for_sm}
\end{figure}

Figure~\ref{fig:figure_3_for_sm} (similar to Fig.~\ref{fig:rates_v_reorg} in the main text) shows our numerical results for three different values of the electronic coupling $V$.

\subsection{Uphill excitonic energy transfer rates}
\begin{figure}[!htbp]
    \centering
    \includegraphics[width=0.75\linewidth]{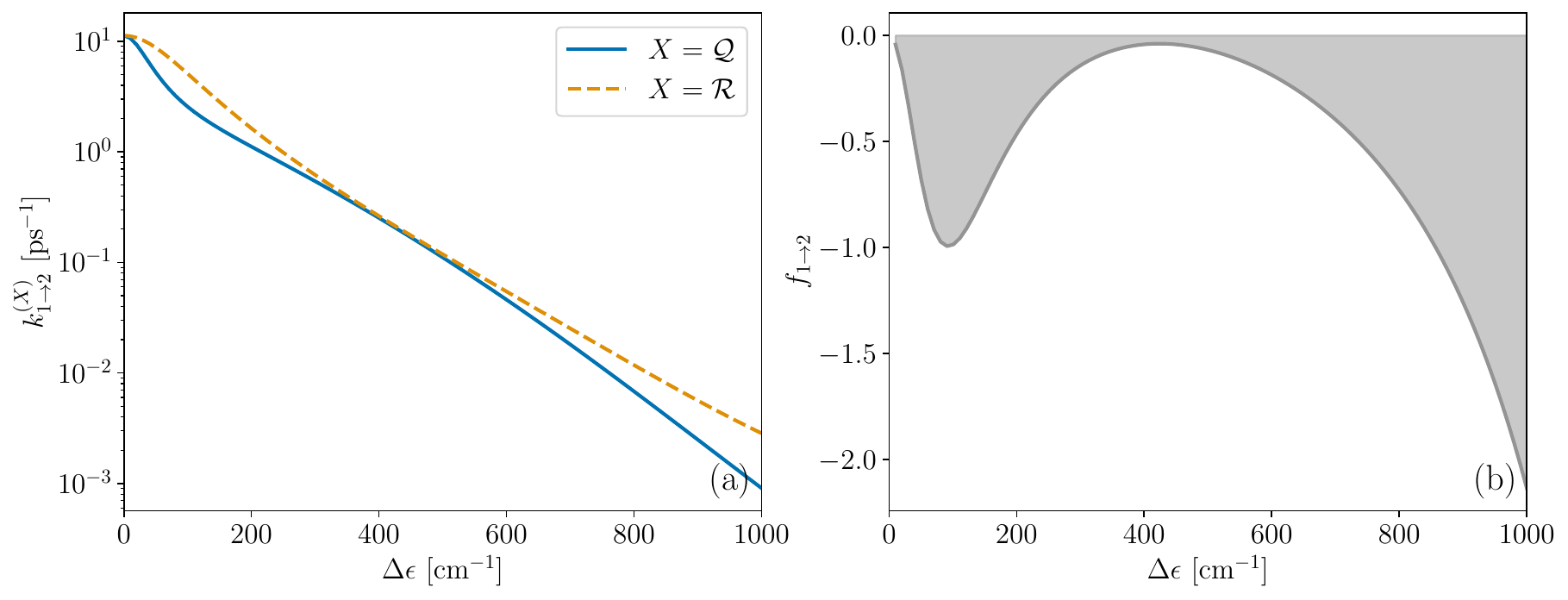}
    \caption{The uphill transfer process results corresponding to Fig.~\ref{fig:rates} in the main text. }
    \label{fig:figure_1_uphill}
\end{figure}

\begin{figure}[!htbp]
    \centering
    \includegraphics[width=0.75\linewidth]{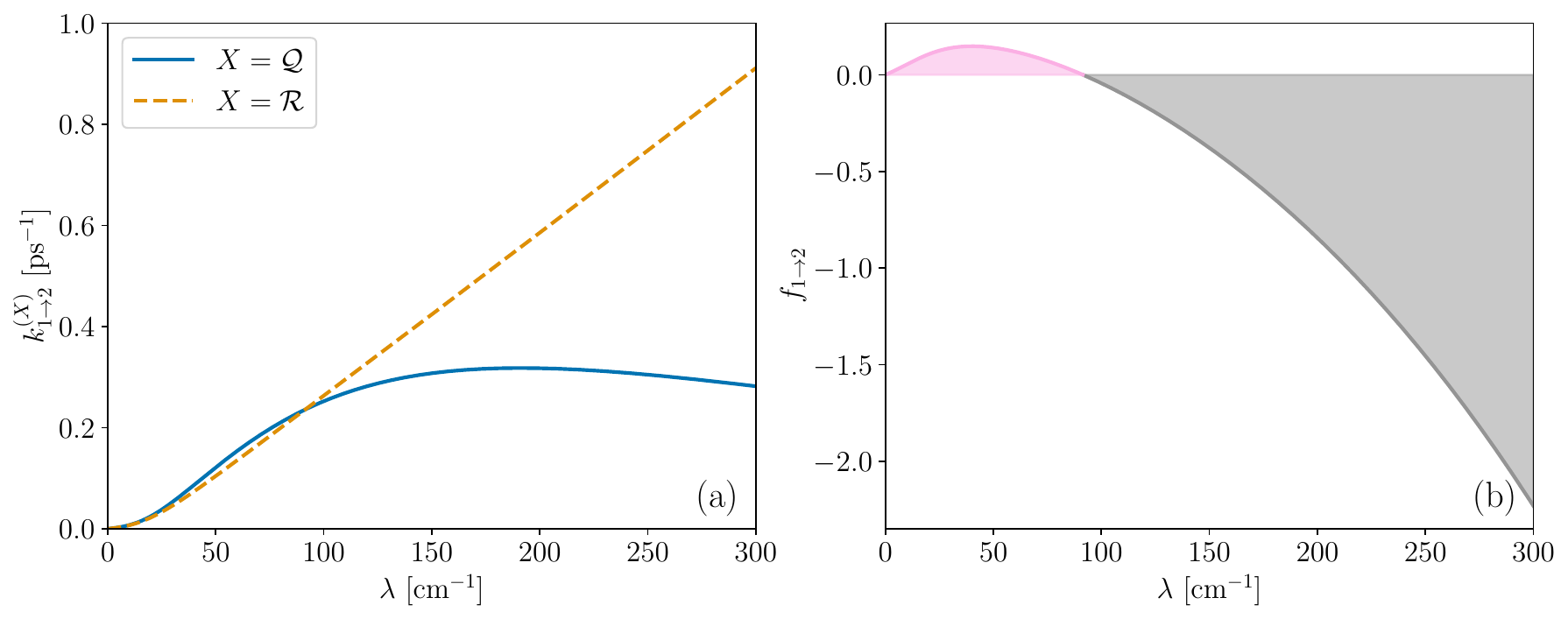}
    \caption{The uphill transfer process results corresponding to Fig.~\ref{fig:rates_v_reorg} in the main text. }
    \label{fig:figure_3_uphill}
\end{figure}

\begin{figure}[!htbp]
    \centering
    \includegraphics[width=0.75\linewidth]{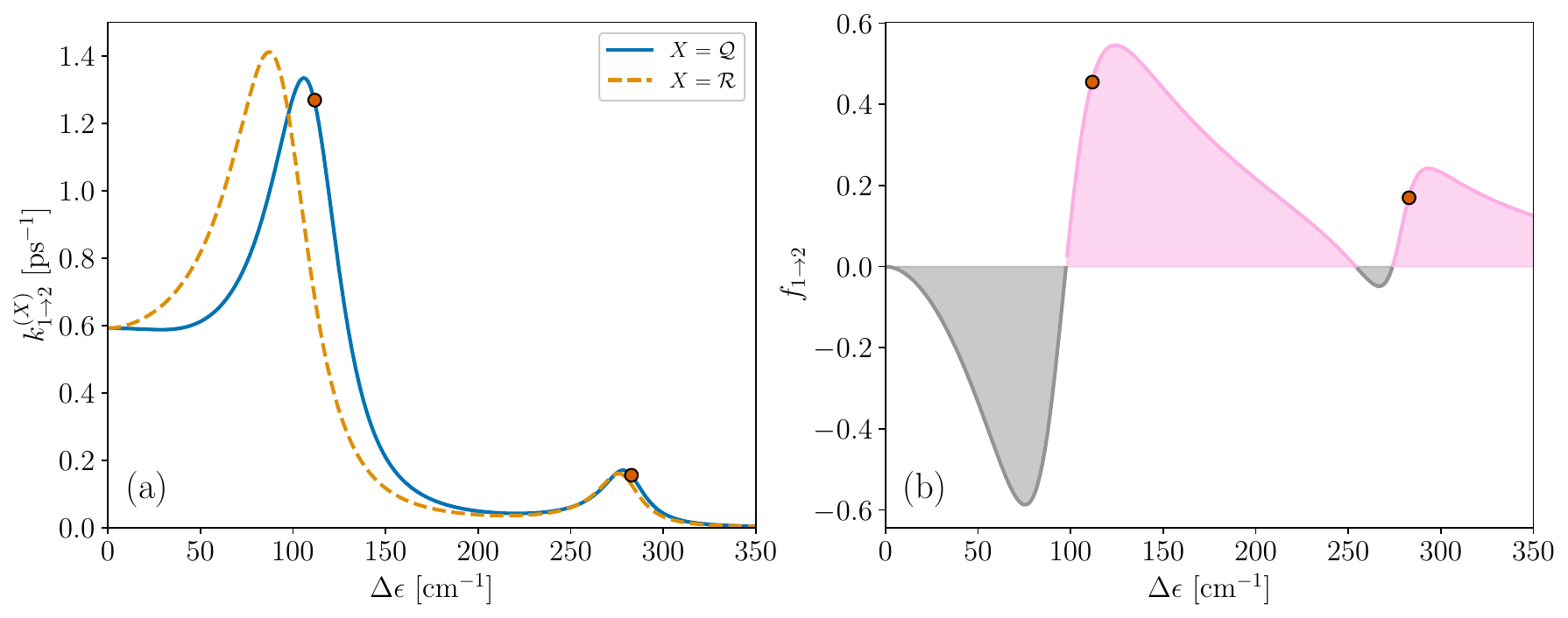}
    \caption{The uphill transfer process results corresponding to Fig.~\ref{fig:rates_with_ubo} in the main text. }
    \label{fig:figure_2_uphill}
\end{figure}

Figures \ref{fig:figure_1_uphill}, \ref{fig:figure_3_uphill}, \ref{fig:figure_2_uphill} show the results of the uphill energy transfer process that correspond to the downhill results we have discussed in the main text.

\subsection{Bare detailed balance}
\begin{figure}[htbp]
    \centering
    \includegraphics[width=0.5\linewidth]{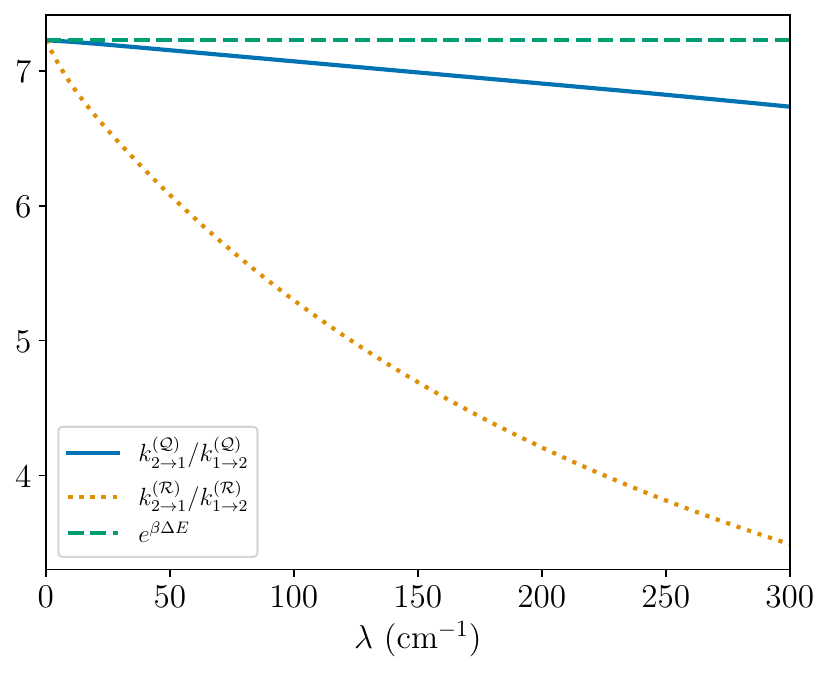}
    \caption{Here we compare the bare detailed balance condition to the numerically converged transfer rates of the dimer model corresponding to Fig.~\ref{fig:rates_v_reorg} in the main text. 
    We have used a truncation depth $N = 21$ and $M = 7$ Matsubara modes. }
    \label{fig:figure_3_detailed_balance}
\end{figure}

Figure~\ref{fig:figure_3_detailed_balance} illustrates how bare detailed balance is broken in general and that it is recovered in the weak-coupling limit $\lambda\to0$. 
The fact that detailed balance does not hold is due to our accurate numerical modeling of the non-Markovian influence of the environment using HEOM.

\end{document}